\documentclass[12pt]{report}
\usepackage[utf8x]{inputenc}
\usepackage{graphicx}
\usepackage{amssymb}
\usepackage[T1]{fontenc}
\usepackage{gfsartemisia}
\usepackage[usenames,dvipsnames]{color}
\usepackage[colorlinks=true, pdfstartview=FitV,linkcolor=PineGreen,citecolor=Bittersweet,urlcolor=Cerulean]{hyperref}

\begin{document}

\thispagestyle{empty}
\begin{center}
{\Large \bf Invariancia de norma, cuantizaci\'on e integraci\'on de modos pesados en una teor\'ia de Kaluza--Klein de norma}
\\
\vspace{0.1cm}
({\large Gauge invariance, quantization and integration of heavy modes in a gauge Kaluza--Klein theory})
\end{center}

\begin{center}
\line(1,0){400}
\end{center}
\vspace{0.5cm}
\begin{center}
{\Large H\'ector Novales S\'anchez}
\end{center}
\vspace{0.5cm}
\begin{figure}[!htb]
\centering
\includegraphics[scale=0.2]{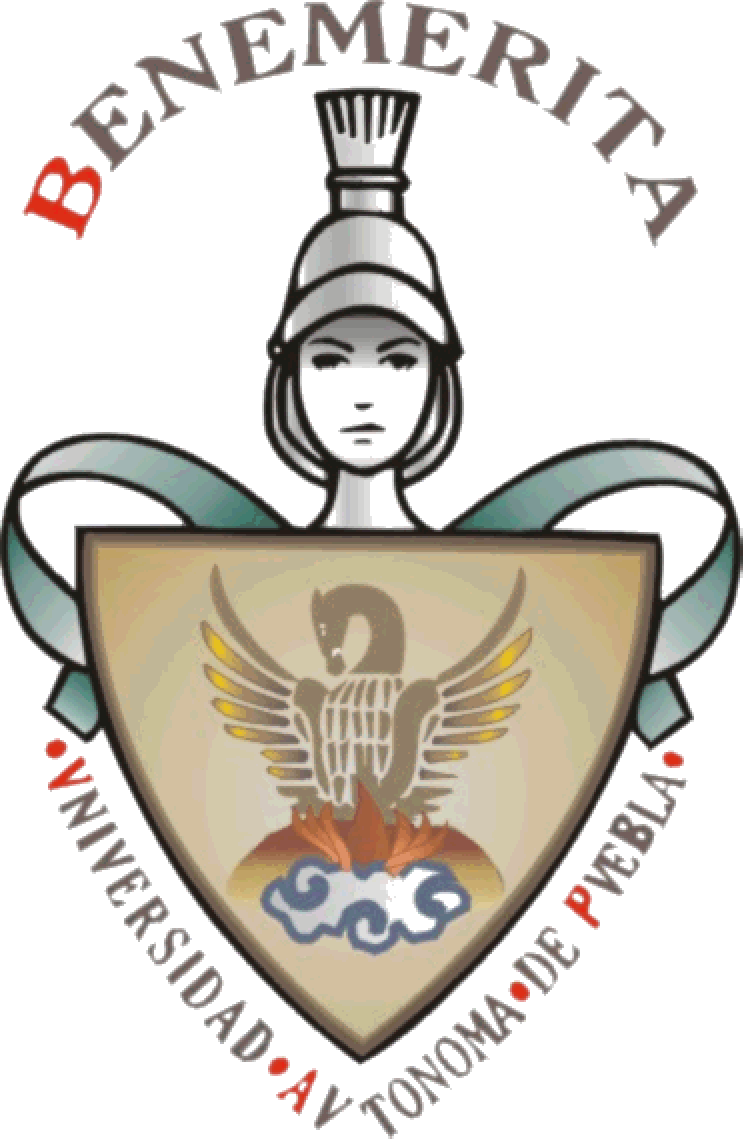}
\end{figure}
\vspace{0.5cm}
\begin{center}
{\large Tesis presentada para obtener el grado de Doctor en Ciencias (F\'isica Aplicada)}
\end{center}
\vspace{0.04cm}
\begin{center}
{\large Supervisado por J. Jes\'us Toscano Ch\'avez}
\end{center}
\vspace{0.04cm}
\begin{center}
{\large Facultad de Ciencias F\'isico Matem\'aticas}
\end{center}
\vspace{0.07cm}
\begin{center}
{\large Benem\'erita Universidad Aut\'onoma de Puebla}
\end{center}

\vspace{1cm}
\begin{center}
Junio de 2012
\end{center}

\newpage
\thispagestyle{empty}
\noindent
{\footnotesize
T\'itulo: Invariancia de norma, cuantizaci\'on e integraci\'on de modos pesados en una teor\'ia de Kaluza--Klein de norma.
\\
Estudiante: M. C. H\'ector Novales S\'anchez
\\
\begin{center}
Comit\'e
\\ \vspace{1.4cm}
\line(1,0){200}
\\ 
Dr. Alfonso Rosado S\'anchez
\\
Presidente
\\ \vspace{1.4cm}
\line(1,0){200}
\\
Dr. Gilberto Tavares Velasco
\\
Secretario
\\ \vspace{1.4cm}
\line(1,0){200}
\\
Dr. J. Lorenzo D\'iaz Cruz
\\
Vocal
\\ \vspace{1.4cm}
\line(1,0){200}
\\
Dr. Alfredo Aranda Fern\'andez
\\
Vocal
\\ \vspace{1.4cm}
\line(1,0){200}
\\
Dr. Miguel \'Angel P\'erez Ang\'on
\\
Vocal
\\ \vspace{1.4cm}
\line(1,0){200}
\\
Dr. Eleazar Cuautle Flores
\\
Suplente
\\ \vspace{1.4cm}
\line(1,0){200}
\\
Dr. J. Jes\'us Toscano Ch\'avez
\\
Asesor
\end{center}
}

\newpage
\thispagestyle{empty}
\section*{Agradecimientos}
Son dos las personas a quienes deseo agradecer de manera especial, pues su invaluable apoyo contribuy\'o determinantemente a la elaboraci\'on de este trabajo de tesis. Primeramente quiero reconocer a mi madre, Mar\'ia de Lourdes S\'anchez de los Santos, a quien debo todo lo que soy. Han sido su apoyo, su comprensi\'on, su dedicaci\'on, su ejemplo, y, sobre todo, su ingente amor factores que han conspirado para fortalecerme a trav\'es de bases s\'olidas y as\'i generar este logro tan importante. Es ella la poseedora de mi m\'as grande y sincera gratitud, y considero que este trabajo y todas sus implicaciones son tambi\'en un logro de ella. Por otra parte, quiero compartir este \'exito con mi asesor, el Dr. J. Jes\'us Toscano Ch\'avez, a quien profeso gran respeto y admiraci\'on. Su apoyo, profesionalismo, entusiasmo y ambici\'on, durante los diferentes niveles de mi formaci\'on profesional, me permitieron alcanzar esta meta. Su gu\'ia, indudablemente, ha marcado profundamente mi vida profesional. Adicionalmente quiero agradecer al Consejo Nacional de Ciencia y Tecnolog\'ia por proporcionarme becas para realizar mis estudios de Maestr\'ia y Doctorado en Ciencias.
\newpage
\thispagestyle{empty}

\section*{Abstract}
\begin{changemargin}{0.7in}{0in}
We start from a pure Yang--Mills theory defined on a spacetime with one universal extra dimension that we compactify on the orbifold $S^1/Z_2$. We obtain a Kaluza--Klein (KK) theory by expanding in KK towers covariant objects rather than fields, as such an approach yields a four--dimensional description possessing an interesting gauge structure in which two sorts of gauge transformations leave, independently of each other, the theory invariant. One type of such transformations are the standard gauge transformations (SGT), which are defined by the zero modes of the gauge parameters, $\alpha^{(0)a}$, and under which the KK zero modes, $A^{(0)a}_\mu$, behave as gauge fields. The other transformations receive the name of nonstandard gauge transformations (NSGT), and under them the KK excited modes, $A^{(m)a}_\mu$, are gauge fields. We then quantize the KK excited modes within the Becchi--Rouet--Stora--Tyutin (BRST) approach, which includes the elimination of the gauge symmetries associated to the KK excitations through a gauge--fixing (GF) procedure that preserves gauge invariance with respect to the SGT. We also present the most general Faddeev--Popov ghost (FPG) sector, which emerges from the BRST quantization process. As a next step, we integrate out the KK excited modes and derive an effective Lagrangian containing the explicit expressions of the coefficients multiplying all the independent nonrenormalizable operators of canonical dimension six that are allowed by the ${\rm SU}_4(N)$ gauge group and by Lorentz invariance. We first perform the calculation in the Feynman--`t Hooft (FtH) gauge and then in the general $R_\xi$ gauge. We find for the latter case a gauge--dependent result. By taking the FtH gauge in such gauge--dependent expression, we consistently recover the result obtained by considering the FtH gauge from the beginning. The derivation of the effective Lagrangian explicitly proves that the contributions of KK excited modes to one--loop light Green's functions are renormalizable. Finally, we compare, at the four--dimensional level, the effects of the extra dimension with the contributions of a presumed fundamental theory describing nature at energies higher than those corresponding to the extra--dimensional physics. We find that the effects of the KK excited modes are the dominant ones.
\end{changemargin}

\tableofcontents
\setcounter{page}{1}

\chapter{Introduction}

The Standard Model (SM) of particle physics~\cite{GSW} is the current best quantum--relativistic theory describing the electromagnetic, weak and strong interactions. It is an elegant model built in the context of quantum field theory, with symmetries playing a major role in its formulation. Despite the fact that the SM has survived to high--precision experiments during many years, the high energy physics community is nowadays convinced that the SM is not the most fundamental theory depicting nature, as there exist physical phenomena which do not find an explanation within it. This is the case, for instance, of gravitational interaction, massive neutrinos, and the experimental evidence that our universe is mostly consituted by some exotic substances known as dark matter and dark energy. Moreover, renormalizability of the SM~\cite{tH}, which is in important issue if one assumes that the SM is all there is, crucialy depends~\cite{RC} on the existence of a scalar particle, the Higgs boson, which has not been observed in nature so far. By considering the SM as a good approximation, valid up to our current experimental sensitivity, the pursuit of a theory of everything can be achieved in mainly two ways. One consists in proposing an ambitious model, which is supposed to be fundamental, and then trying to connect it, under certain circumstances, with the SM, as it occurs, for instance, in the case of string theory. Of course, the lack of experimental evidence for very high energy phenomena makes this path quite difficult and much work has to be done on the theoretical side. Another, more economical, track that could shed light on the question of the most fundamental theory is accompished by establishing {\it SM extensions} and see if they are capable to explain some issues out of the reach of our current low--energy picture. If a given SM extension reconciles experiments with theory, then some clues on the fundamental theory should arise.

There are interesting SM extensions in the market, and those that rely on the existence of extra dimensions are presently among the most popular ones. Extra--dimensional models became phenomenologically interesting, some years ago, when it was realized that the size of extra dimensions could be in the TeV$^{-1}$ range~\cite{AADD}. Since then, a plethora of works concerning extra dimensions has appeared, leading to interesting physical implications such as a lower gauge--couplings unification scale~\cite{DDG1,DDG2}, in the supersymmetric context, and a solution to the hierarchy problem~\cite{RS}. One of the main motivations of extra--dimensional models is string theory, which supposes that the full spacetime, which is known as the {\it bulk}, has eleven spacetime dimensions where our three--dimensional space, usually called {\it 3--brane}, is embedded. The easiest way to propound an extra--dimensional model is taking the already known fields and define all or some of them to propagate in the bulk. The scenario in which {\it all} of the fields propagate in the extra dimensions is known as the {\it universal extra dimensions}~\cite{UED1} (UED) framework. With respect to the interesting new physics generated by UED models, it is worth commenting that they provide~\cite{STandCFMandH} dark matter candidates. These models have been also employed to develope new mechanisms for generating~\cite{ADPY1} neutrino masses and supressing~\cite{ADPY2} proton decay.

Thinking about extra dimensions, one could ask: what is the character of them? Currently, there is no experimental evidence indicating that they actually exist, so there is plenty of room to build extra--dimensional models with different features. Models with {\it spatial} extra dimensions are the ones favored in the literature~\footnote{For a discussion on time-like extra dimensions, see~\cite{timeED}.}. The compactification of extra dimensions is a crucial point of extra dimensional models, for it is an experimentally consistent way to explain why we have not detected them so far. This is achieved by assuming that the extra dimensions do not extend infinitely, but are {\it compactified} and their size is small enough to render them invisible to present--day experiments. The propagation of particles in such finite dimensions furnishes the corresponding extra--dimensional fields with periodicity properties that allow one to expand the fields in Fourier series with respect to the extra--dimensional coordinates. Such series, also known as {\it Kaluza--Klein} (KK) {\it towers}, involve infinite sets of fields that receive the name of {\it KK modes} and that can be divided into two types: the {\it zero modes}, which are the fields appearing in the first term of each KK tower, and which are identified with the SM fields; and the {\it excited modes}, which are associated to the other terms in the Fourier series and correspond to new four--dimensional particles. The confinement of fields within the finite extra dimensions quantizes the allowed states, so that the masses of the KK modes are highly degenerated. This is analogous to the case of the infinite--well potential that is studied in elementary quantum mechanics~\cite{TRrev}. The compactification of extra dimensions also introduces a parameter to the theory, namely, the size of the extra dimensions, so that all physical quantities built from extra--dimensional models are expected to depend on it.

In this thesis work, we consider an ${\rm SU}_5(N)$--invariant Yang--Mills theory defined in a spacetime manifold with one flat UED, which we compactify on the orbifold $S^1/Z_2$ with radius $R$. The Yang--Mills theory~\cite{YMtheory} has great physical interest, for it is used to suitably describe the strong and electroweak interactions with the complicity of symmetry. Hence, investigating its five--dimensional generalization and the physical implications generated by it on low--energy physics is, of course, worthy. Through the rest of this paper, the ordinary four--dimensional coordinates will be labeled by $x$ and the extra dimension will be represented by $y$. Capital letters indices will run over the five components of the five--dimensional spacetime ($M,N,\ldots=0,1,2,3,5$), while greek indices will take values corresponding to the ordinary four--dimensional coordinates ($\mu,\nu,\ldots=0,1,2,3$). Gauge group indices will be represented by lower--case letters ($a,b\ldots=1,2,\ldots,N^2-1$). The five--dimensional Yang--Mills Lagrangian is defined as
\begin{equation}
{\cal L}_{\rm YM}^{\rm 5\,dim}(x,y)=-\frac{1}{4}{\cal F}^a_{MN}(x,y){\cal F}^{aMN}(x,y),
\label{L5YM}
\end{equation}
where the curvatures ${\cal F}^a_{MN}$ are given by
\begin{equation}
{\cal F}^a_{MN}(x,y)=\partial_M{\cal A}^a_N(x,y)-\partial_N{\cal A}^a_M(x,y)+g_5f^{abc}{\cal A}^b_M(x,y){\cal A}^c_N(x,y).
\end{equation}
In the last expression, the five--dimensional gauge fields were denoted by ${\cal A}^a_M(x,y)$, $g_5$ is the five--dimensional coupling constant with units of $({\rm mass})^{-1/2}$ and $f^{abc}$ are the structure constants associated with the ${\rm SU}_5(N)$ gauge group. It is well known that extra--dimensional theories are not renormalizable, for they involve coupling constants with inverse--mass dimensions. This indicates that the extra--dimensional Yang--Mills theory does not incarnate a fundamental theory, but only a formulation valid up to certain energy scale, $M_{\rm S}$, above which nature behaves according to a fundamental description, which could be, perhaps, string theory. In fact, the nonrenormalizable behavior of this description implies that the extra--dimensional theory is not even valid for an arbitrarily large energy range. In connection with that, an analysis of the running of the coupling constants, within the context of the SM with one UED, has shown~\cite{UED1,UED2} that $M_{\rm S}R\sim 30$. The effects of the fundamental theory within the range of energies between the compactification scale $R^{-1}$ and the physical cutoff $M_{\rm S}$ can be parametrized through an effective Lagrangian expansion. As the five--dimensional theory is nonrenormalizable, there is no limit for the number of invariants that can be included in the Lagrangian. This has been utilized, for instance, to introduce~\cite{Rlv} Lorentz violation in extra dimensions in the same way that it is done in the SM Extension~\cite{Ksme}. In the present work we do not pay attention to such effects. The full effective theory can be written as~\cite{NT1,NT2}
\begin{equation}
{\cal L}_{\rm eff}^{\rm 5\,dim}={\cal L}_{\rm YM}^{\rm 5\,dim}+\sum_{N}\beta_N\frac{g_5^{n_N}}{M_{\rm S}^{m_N}}{\cal O}_N^{\rm 5\,dim}({\cal A}^a_M),
\label{full5Deff}
\end{equation}
where the ${\cal O}^{\rm 5\,dim}_N$ are operators of canonical dimension higher than five that are invariant under the five--dimensional Lorentz group as well as under the ${\rm SU}_5(N)$ gauge transformations. The $\beta_N$ coefficients are dimensionless factors that quantify the effects of the fundamental theory at the extra--dimensional level provided one considers energies below the $M_{\rm S}$ scale. The mass dimension of each term is regulated by appropriate powers of the dimensionful coupling constant $g_5$ and the inverse of the fundamental scale $M_{\rm S}$. The suppression introduced by the fundamental scale is important because it renders~\cite{UED1,NT2,FMNRT} these effects dominated by those of the extra dimensions. By orbifold--compactifying the extra dimension, the effective theory described by Eq.(\ref{full5Deff}) produces the KK theory
\begin{equation}
{\cal L}_{\rm eff}^{\rm 4\,dim}=\int_0^{\pi R}dy\, {\cal L}_{\rm eff}^{\rm 5\,dim}={\cal L}_{\rm YM}^{\rm 4\,dim}(\phi^{(0)},\phi^{(n)})+\sum_{N}\alpha_N\frac{R^{j_N}}{M^{k_N}_{\rm S}}{\cal O}_N^{\rm 4\,dim}(\phi^{(0)},\phi^{(n)}),
\label{Leff4}
\end{equation}
with $\phi^{(0)}$ and $\phi^{(n)}$ generically and respectively representing all the zero modes and excited modes, and where
\begin{equation}
{\cal L}_{\rm YM}^{\rm 4\,dim}=\int_0^{\pi R}dy\,{\cal L}_{\rm YM}^{\rm 5\,dim}.
\end{equation}
As it has been indicated in the last expression, the ${\cal L}_{\rm YM}^{\rm 4\,dim}$ term, whose dynamic variables are the KK modes, was produced by the five--dimensional Yang--Mills Lagrangian after integrating out the fifth dimension. The appropriate expansion of covariant objects (the curvatures) in Eq.(\ref{L5YM}) implies\footnote{The KK expansion of fields instead of covariant objects leads to a completely different four--dimensional KK theory. For further discussion on this issue, see Refs.~\cite{NT1,NT3,NT4}.} that ${\cal L}_{\rm YM}^{\rm 4\,dim}$ is invariant  under two types of infinitesimal gauge transformations~\cite{NT1}. The first set of such variations are the {\it standard gauge transformations}~\cite{NT1} (SGT), under which the zero modes transform as gauge fields. The other sort of local gauge transformations receive the name of {\it nonstandard gauge transformations}~\cite{NT1}, and the fields transforming as gauge fields under them are KK excited modes. The second term in the right--hand side of Eq. (\ref{Leff4}) is a sum of terms containing nonrenormalizable operators constituted by KK modes, both zero and excited ones. As the mass dimension of such objects is higher than four, the factors $R^{j_1}/M^{k_2}_{\rm S}$ control the total canonical dimension of the terms. The dimensionless $\alpha_N$ coefficients parametrize the effects of the physical description of nature beyond the $M_{\rm S}$ scale at the low--energy level, at which the KK modes are the dynamic variables.

The quantization of gauge systems requires the fixation of the gauge, for leaving such degeneration would render the path integral divergent, since a set of physically equivalent trayectories, related to each other by gauge transformations, whould be taken into account. After the explicit breaking of gauge symmetry through the election of a particular gauge, what remains is the so--called {\it Becchi--Rouet--Stora--Tyutin} (BRST) symmetry~\cite{BRST}. This is a global symmetry that naturally arises within the field--antifield formalism~\cite{GPS}. The quantization within the BRST approach comprehends the gauge--fixing (GF) procedure, which enters in a nontrivial way and comes along with the consistent determination of the most general Faddeev--Popov ghost (FPG) sector. As the KK theory emerged from the five--dimensional Yang--Mills Lagrangian is a gauge theory, the BRST approach is a suitable tool to consistently quantize it. Once this step has been performed, two terms are added to the KK Lagrangian ${\cal L}_{\rm YM}^{\rm 4\,dim}$, namely, the {\it gauge--fixing} Lagrangian, ${\cal L}_{\rm GF}$, and the {\it Faddeev--Popov ghost} term, ${\cal L}_{\rm FPG}$. As the SGT are defined exclusively by the zero modes of the gauge parameters, while the NSGT are determined only by the KK excitations of such parameters,
the GF procedure for this theory can be split into two independent parts. One can fix the gauge for the KK excited modes and leave the gauge degeneracy associated to the KK zero modes. After that one can, if desired, fix the gauge for the remaining gauge fields. Calculation of extra--dimensional effects on low--energy physics requires only to quantize the KK excited modes and consider the zero modes as classical fields. In relation with that, it is worth emphasizing that the first corrections to SM observables from extra dimensions, in the UED framework, enter~\cite{UED1} since the one--loop level, while no tree--level corrections to such quantities exist. This is a consequence of so--called KK number conservation~\cite{UED1}, which is characteristic of UED models. The effective Lagrangian then reads
\begin{equation}
{\cal L}_{\rm eff}^\xi={\cal L}_{\rm YM}^{\rm 4\,dim}(\phi^{(0)})+{\cal L}_{\rm GF}(\phi^{(0)},\phi^{(n)};\xi)
+{\cal L}_{\rm FPG}(\phi^{(0)},\phi^{(n)};\xi)
+\sum_{N}\alpha_N\frac{R^{j_N}}{M^{k_N}_{\rm S}}{\cal O}_N^{\rm 4\,dim}(\phi^{(0)},\phi^{(n)}).
\label{Leffcompleto}
\end{equation}
Here, $\xi$ is used to denote the {\it gauge--fixing parameter}, whose different values correspond to different choices of the gauge. The last term in Eq.(\ref{Leffcompleto}) is a sum of nonrenormalizable operators suppressed by the compactification scale and by the fundamental physics scale as well. The powers of these objects are such that the mass dimension of any complete term is four. Each of these terms also includes a dimensionless parameter, $\alpha_N$, that bears quantitative information about the effects of the fundamental description at low energy. As the KK excitations are heavier than the zero--mode fields, the most important terms in this series are those that involve no other fields than the zero modes. For that reason, we disregard, in what follows, those parts of the fourth term of Eq.(\ref{Leffcompleto}) involving KK excited modes. By integrating out the KK excitations in the other terms, one obtains~\cite{NT2} an effective Lagrangian expansion that depends only on zero--mode gauge fields. Schematically, the result of this process is
\begin{equation}
{\cal L}^\xi_{\rm eff}={\cal L}_{\rm YM}(\phi^{(0)})+\sum_{N}\kappa_N(\xi)\,R^{m_N}\,{\cal O}_N^{\rm KK}(\phi^{(0)})+\sum_N\alpha_N\left(\frac{R}{M_{\rm S}}\right)^{r_N}{\cal O}^{\rm 4\,dim}_N(\phi^{(0)}),
\label{Leffintegrated}
\end{equation}
where ${\cal L}_{\rm YM}$ is the ordinary four--dimensional Yang--Mills Lagrangian. The second term is a series constituted by operators of mass dimension higher than four, whose building blocks are low--energy fields (KK zero modes) and low--energy symmetries (four--dimensional Lorentz and ${\rm SU}_4(N)$). This sum of nonrenormalizable invariants is equivalent to a sum of light Green's functions with one--loop quantum corrections introduced by the KK excited modes. As the heavy KK modes that have been integrated out are gauge fields, the coefficients $\kappa_N(\xi)$ muliplying the nonrenormalizable operators parametrize the effects of extra--dimensional physics in a gauge--dependent manner. Of course, it is expected that the GF parameter vanishes when a physical quantity is constructed. Note that the factors of powers of the compactification radius, $R^{m_N}$, supress the extra--dimensional effects on light Green's functions, as the size of the extra dimension is assumed, on the grounds of experimental consistency, to be small. The presence of such factors also shows explicitly that the extra--dimensional contributions decouple for a large compactification scale (or a small compactification radius), which is consistent, for the well known renormalizability of the Yang--Mills theory~\cite{tH} sets~\cite{W} the required conditions for the decoupling theorem~\cite{AC} to be fulfilled. It is worth commenting that the effective Lagrangian ${\cal L}_{\rm eff}^\xi$, Eq.(\ref{Leffintegrated}), permits one to compare~\cite{NT2} the effects of the extra dimension with those from the fundamental description beyond the cutoff $M_{\rm S}$, and the result is~\cite{NT2} that the latter are negligible with respect to the former. This asseveration was already pointed out~\cite{UED1} and, recently, phenomenologically illustrated~\cite{FMNRT} by comparing the one--loop contributions to the $WW\gamma$ and $WWZ$ vertices produced by the extra dimension with those provided by the parametrization of the fundamental physical description at the tree level. This result, Eq.(\ref{Leffintegrated}), also proves that the one--loop level effects from the KK excited modes on light Green's functions are~\cite{NT2} renormalizable, which had been shown before in the literature~\cite{NT1} by following a different path. It occurs that the divergencies originated in the loop integrals are {\it eaten} by the parameters of the low--energy theory and are, consequently, unobservable.

The rest of the paper is organized as follows. The necessary framework to  perform the integration of KK excited modes is established and discussed in Chapter~\ref{5DYMt}. The full discussion includes the derivation of a KK theory, with especial emphasis on the nature of gauge symmetry, and the quantization of the KK excited modes, within the BRST approach. Chapter~\ref{KKint} deals with the integration of the KK excited modes and the determination of an effective Lagrangian comprehending the one--loop effects of these heavy fields on light Green's functions. The calculation is performed first in the Feynman--`t Hooft (FtH) gauge and then in the $R_\xi$ gauge. Also, a comparison among the effects of extra--dimensional physics and physics related to a fundamental description is carried out. Finally, the conclusions are presented in Chapter~\ref{conclusions}.

\chapter{The five--dimensional Yang--Mills theory}
\label{5DYMt}
Consider a five--dimensional spacetime manifold and assume that the extra dimension, which we suppose to be spatial, is flat. Take the five--dimensional Yang--Mills Lagrangian ${\cal L}_{\rm YM}^{\rm 5\,dim}$, which was defined in Eq. (\ref{L5YM}) and whose dynamic variables are gauge vector fields with five components,
\begin{equation}
{\cal A}_M(x,y):({\cal A}_0(x,y),{\cal A}_1(x,y),{\cal A}_2(x,y),{\cal A}_3(x,y),{\cal A}_5(x,y)),
\end{equation}
which we utilize to define the five--dimensional field strengths ${\cal F}^a_{MN}(x,y)$ as usual. This gauge theory is governed by the five--dimensional Lorentz symmetry and by the ${\rm SU}_5(N)$ gauge symmetry as well. According to gauge symmetry, the transformations
\begin{equation}
{\cal A}^a_M(x,y)\to{\cal A}^a_M(x,y)+{\cal D}^{ab}_M\,\alpha^b(x,y),
\end{equation}
with $\alpha^a(x,y)$ representing the gauge parameters and ${\cal D}^{ab}_M=\delta^{ab}\partial_M-g_5f^{abc}{\cal A}^c_M$, vary the ${\cal F}^a_{MN}$ covariantly and leave the ${\cal L}_{\rm YM}^{\rm 5\,dim}$ Lagrangian invariant. From the viewpoint of the BRST formulation, the gauge parameters coincide~\cite{GPS} with the ghost fields and hence personify dynamic variables of the theory, at the same level of the gauge fields. As the extra dimension is universal, the gauge parameters must then propagate in it. This is moreover consistent with five--dimensional Lorentz invariance\footnote{In the case in which the gauge parameters do not depend on the extra dimension, Lorentz symmetry is explicitly violated.}. One could, of course, consider a five--dimensional Yang--Mills theory in which the gauge parameters are constrained to live in our ordinary four--dimensional spacetime, but the physical implications of such framework are less interesting~\cite{NT1,NT3}. For this reason we will not consider such possibility in the present work.

\section{The Kaluza--Klein theory}
As experiments have not observed any signal that suggests the existence of extra dimensions, it is necessary to introduce a mechanism to render this description experimentally consistent, and compactification is a rather elegant way to do it. The main idea is that the extra dimension does not extend infinitely, as those of our known ordinary four--dimensional world, but it is limited to a finite region. Then the size of the extra dimension is assumed to be small enough so that their effects have been so far innocuous to our experiments. A physically suitable scheme, which we will follow to compactify the extra dimension, is the compactification on the orbifold $S^1/Z_2$, which we describe now. We suppose that the extra dimension closes in itself, forming a circle of radius $R$. After that, we impose a $Z_2$ symmetry on the extra dimension, which identifies every point $y$ of the extra dimension with its negative counterpart $-y$. The resulting extra dimension, which is said to be compactified on the orbifold $S^1/Z_2$ with radius $R$, is an interval with two singularities. There is interesting physics concerning such singularities~\cite{CGMPT}, but we will not disscuss this issue in the present paper. The radius $R$, which carries the information of the size of the extra dimension, determines a compactification scale, $R^{-1}$, from which the physics of the extra--dimensional extension enters as the valid physical theory describing nature beyond the standard theory.

The orbifold compactification provides the fields and gauge parameters of the theory with periodicity properties with respect to the extra dimension,
\begin{eqnarray}
{\cal A}^a_M(x,y)&=&{\cal A}^a_M(x,y+2\pi R),
\\
\alpha^a(x,y)&=&\alpha^a(x,y+2\pi R),
\end{eqnarray}
and allows one to endow the gauge fields with the parity properties\footnote{This is not the only possible election for the parity properties of the five--dimensional gauge fields. Indeed, assuming the fifth component ${\cal A}^a_5$ to be even under extra coordinate reflection is the base of the so--called gauge--Higgs unification models~\cite{GHu}.} 
\begin{eqnarray}
\label{pmu}
{\cal A}^a_\mu(x,y)&=&{\cal A}^a_\mu(x,-y),
\\
\label{p5}
{\cal A}^a_5(x,y)&=&-{\cal A}^a_5(x,-y),
\end{eqnarray}
under reflection of the $y$ coordinate. Note that the Lagrangian ${\cal L}_{\rm YM}^{\rm 5\,dim}(x,y)$ is invariant under these parity transformations. On the other hand, the behavior of the five--dimensional gauge parameters under parity can be derived~\cite{MPR} from the parity transformation properties of the gauge fields by appealing to five--dimensional gauge symmetry, which leads to the conclusion that
\begin{equation}
\alpha^a(x,y)=\alpha^a(x,-y).
\end{equation}
The dynamic variables of the theory can then be expanded in Fourier series as
\begin{eqnarray}
\label{KKAmu}
{\cal A}^a_\mu(x,y)&=&\frac{1}{\sqrt{\pi R}}A^{(0)a}_\mu(x)+\sum_{n=1}^\infty\sqrt{\frac{2}{\pi R}}A^{(n)a}_\mu(x)\,{\rm cos}\left( \frac{ny}{R} \right),
\\ \nonumber \\
\label{KKA5}
{\cal A}^a_5(x,y)&=&\sum_{n=1}^\infty\sqrt{\frac{2}{\pi R}}A^{(n)a}_5(x)\,{\rm sin}\left( \frac{ny}{R} \right),
\\ \nonumber \\
\label{KKgp}
\alpha^a(x,y)&=&\frac{1}{\sqrt{\pi R}}\alpha^{(0)a}(x)+\sum_{n=1}^\infty\sqrt{\frac{2}{\pi R}}\alpha^{(n)a}(x)\,{\rm cos}\left( \frac{ny}{R} \right),
\end{eqnarray}
where the indices between parentheses are mode numbers labeling the terms of the expansions. These expansions are known, in the high energy physics terminology, as KK towers. The most remarkable feature of the KK towers is the presence of an infinite set of fields that receive the name of KK modes. The KK modes depend, exclusively, on the four--dimensional spacetime coordinates, while all the dependence on the fifth coordinate is located in the arguments of the trigonometric functions of the expansions.  The fields in the first term of each of the KK towers are known as {\it zero modes} and are identified as the low--energy fields. For instance, in the context of the SM colour group these light fields correspond to the known SM gluons. The fields in the rest of the terms of the expansions are called {\it excited modes}, and they represent new particles predicted by the theory. At energies not too far beyond the compactification scale, the dynamic variables of the theory are the KK modes, but as one explores higher energies, the compactification of the extra dimension should become unnoticeable and the dynamic variables should be the five--dimensional gauge fields. A profound feature of the KK towers of the gauge parameters is that such Fourier decompositions engender an infinite number of gauge parameters propagating in the four--dimensional spacetime. As each of these gauge parameters should define a gauge transformation, an infinite set of gauge transformations is expected to manifest. This means that the compactification of the extra dimension comes along with a severe modification of the gauge structure of the extra--dimensional theory. This assertion can be better understood by recalling that gauge and Lorentz symmetries live strongly connected to each other in a nontrivial way. The compactification procedure implies the election of a prefered direction in the five--dimensional spacetime and hence produces an explicit breaking of Lorentz symmetry. Such breaking deeply affects gauge symmetry through its symbiotic link with Lorentz symmetry. 

In the context of UED, the objects to KK expand are~\cite{NT1,NT3} the covariant objects ${\cal F}^a_{MN}(x,y)$ instead of the fields ${\cal A}^a_M(x,y)$, as such a procedure preserves~\cite{NT1,NT2} enough gauge symmetry to consistently produce a gauge transformation per each KK mode of the gauge parameters. The parity and periodicity properties of the five--dimensional gauge fields imply analogous transformation properties of the field strenghts, that is,
\begin{eqnarray}
{\cal F}^a_{\mu\nu}(x,y)={\cal F}^a_{\mu\nu}(x,y+2\pi R), && {\cal F}^a_{\mu\nu}(x,y)={\cal F}^a_{\mu\nu}(x,-y),
\\ \nonumber \\
{\cal F}^a_{\mu5}(x,y)={\cal F}^a_{\mu5}(x,y+2\pi R),&&{\cal F}^a_{\mu5}(x,y)=-{\cal F}^a_{\mu5}(x,-y).
\end{eqnarray}
Such properties allow one to KK--expand these five--dimensional curvatures as
\begin{eqnarray}
\label{KKFmunu}
{\cal F}^a_{\mu\nu}(x,y)&=&\frac{1}{\sqrt{\pi R}}{\cal F}^{(0)a}_{\mu\nu}(x)+\sum_{n=1}^\infty\sqrt{\frac{2}{\pi R}}{\cal F}^{(n)a}_{\mu\nu}(x)\,{\rm cos}\left( \frac{ny}{R} \right),
\\ \nonumber \\
\label{KKFmu5}
{\cal F}^a_{\mu5}(x,y)&=&\sum_{n=1}^\infty\sqrt{\frac{2}{\pi R}}{\cal F}^{(n)a}_{\mu5}(x)\,{\rm sin}\left( \frac{ny}{R} \right).
\end{eqnarray}
As the dependence on the extra--dimensional coordinate is totally embedded in trigonometric functions, the extra dimension can be trivially integrated out in the action,
\begin{equation}
S_0=\int d^4x\int_0^{\pi R} dy\;{\cal L}_{\rm YM}^{\rm 5\,dim}(x,y)\equiv\int d^4x\;{\cal L}_{\rm YM}^{\rm 4\,dim}(x),
\end{equation}
which produces the four--dimensional KK effective Lagrangian ${\cal L}_{\rm YM}^{\rm 4\,dim}(x)$. The precise form of the KK Lagrangian ${\cal L}_{\rm YM}^{\rm 4\,dim}$ is~\cite{NT1}
\begin{equation}
{\cal L}_{\rm YM}^{\rm 4\,dim}=-\frac{1}{4}\left( {\cal F}^{(0)a}_{\mu\nu}{\cal F}^{(0)a\mu\nu}+{\cal F}^{(n)a}_{\mu\nu}{\cal F}^{(n)a\mu\nu}+2{\cal F}^{(n)a}_{\mu5}{\cal F}^{(n)a\mu 5} \right),
\end{equation}
where any pair of repeated indices, including the modes ones, indicates a sum. Notice that the well--defined parity of the five--dimensional field strengths and the orthogonality of trigonometric functions ensure that no mixings among different KK modes of curvatures appear. The expressions of the KK curvatures in terms of the KK modes of the five--dimensional vector bosons can be straightforwardly determined~\cite{NT1} by employing the parity and orthogonality properties of trigonometric functions. The recipe is simple: KK--expand the components of the five--dimensional curvature, ${\cal F}^a_{\mu\nu}$ and ${\cal F}^a_{\mu5}$, in terms of the KK vector bosons  by utilizing Eqs.(\ref{KKAmu}) and (\ref{KKA5}); equalize the resulting expressions to Eqs. (\ref{KKFmunu}) and (\ref{KKFmu5}) and obtain
\begin{equation}
\label{EqF1}
\begin{array}{lcl}
\displaystyle
\frac{1}{\sqrt{\pi R}}F^{a}_{\mu\nu}
\\ \\
\displaystyle
+\sum_{m=1}^\infty\sqrt{\frac{2}{\pi R}}\left( D^{ab}_\mu A^{(m)b}_\nu-D^{ab}_\nu A^{(m)b}_\mu \right){\rm cos}\left( \frac{my}{R} \right)
\\ \\
\displaystyle
+gf^{abc}\sum_{m=1}^\infty\sum_{n=1}^\infty\frac{2}{\sqrt{\pi R}}A^{(m)b}_\mu A^{(n)c}_\nu {\rm cos}\left( \frac{my}{R} \right){\rm cos}\left( \frac{ny}{R} \right)&=&\displaystyle\frac{1}{\sqrt{\pi R}}{\cal F}^{(0)a}_{\mu\nu}
\\ \\&&
\displaystyle+\sum_{m=1}^\infty\sqrt{\frac{2}{\pi R}}{\cal F}^{(n)a}_{\mu\nu}{\rm cos}\left( \frac{my}{R} \right),
\end{array}
\end{equation}
\begin{equation}
\label{EqF2}
\begin{array}{lcl}
\displaystyle\sum_{m=1}^\infty\sqrt{\frac{2}{\pi R}}\left(  D^{ab}_\mu A^{(m)b}_5+\frac{m}{R}A^{(m)a}_\mu \right){\rm sin}\left( \frac{my}{R} \right)
\\ \\
\displaystyle+gf^{abc}\sum_{m=1}^\infty\sum_{n=1}^\infty\frac{2}{\sqrt{\pi R}}A^{(m)b}_\mu A^{(n)c}_5{\rm cos}\left( \frac{my}{R} \right){\rm sin}\left( \frac{ny}{R} \right)&=&\displaystyle\sum_{m=1}^\infty\sqrt{\frac{2}{\pi R}}{\cal F}^{(m)a}_{\mu5}{\rm sin}\left( \frac{my}{R} \right),
\end{array}
\end{equation}
where
\begin{eqnarray}
F^{a}_{\mu\nu}&=&\partial_\mu A^{(0)a}_\nu-\partial_\nu A^{(0)a}_\mu+gf^{abc}A^{(0)b}_\mu A^{(0)c}_\nu,
\\ \nonumber \\
D^{ab}_\mu&=&\delta^{ab}\partial_\mu-gf^{abc}A^{(0)c}_\mu,
\end{eqnarray}
are, respectively, the four--dimensional Yang--Mills field strength and the covariant derivative in the adjoint representation of the ${\rm SU}_4(N)$ group, and $g=g_5/\sqrt{\pi R}$ is the dimensionless coupling constant of the four--dimensional Yang--Mills theory; integrate both sides of Eq.(\ref{EqF1}) on the interval $0$ to $\pi R$ to derive ${\cal F}^{(0)a}_{\mu\nu}$; multiply both sides of Eq.(\ref{EqF1}) by ${\rm cos}(ky/R)$ and integrate from $0$ to $\pi R$ to find the expression of ${\cal F}^{(m)a}_{\mu\nu}$; finally, multiply the two sides of Eq.(\ref{EqF2}) by ${\rm sin}(ky/R)$, integrate over $0$ to $\pi R$ and obtain ${\cal F}^{(m)a}_{\mu5}$. The resulting expressions for the curvatures are~\cite{NT1}
\begin{eqnarray}
{\cal F}^{(0)a}_{\mu\nu}&=&F^{a}_{\mu\nu}+gf^{abc}A^{(m)b}_\mu A^{(m)c}_\nu,
\\ \nonumber \\
{\cal F}^{(m)a}_{\mu\nu}&=&D^{ab}_\mu A^{(m)b}_\nu-D^{ab}_\nu A^{(m)b}_\mu+gf^{abc}\Delta^{mrn}A^{(r)b}_\mu A^{(n)c}_\nu,
\\ \nonumber \\
{\cal F}^{(m)a}_{\mu5}&=&D^{ab}_\mu A^{(m)b}_5+\frac{m}{R}A^{(m)a}_\mu+gf^{abc}\Delta'^{mrn}A^{(r)b}_\mu A^{(n)c}_5,
\end{eqnarray}
with
\begin{eqnarray}
\Delta^{mrn}&=&\frac{1}{\sqrt{2}}\left( \delta^{r,m+n}+\delta^{m,r+n}+\delta^{n,r+m} \right),
\\ \nonumber \\
\Delta'^{mrn}&=&\frac{1}{\sqrt{2}}\left( \delta^{r,m+n}+\delta^{m,r+n}-\delta^{n,r+m} \right).
\end{eqnarray}
Note that the zero--mode curvature ${\cal F}^{(0)a}_{\mu\nu}$ includes the ordinary four--dimensional Yang--Mills curvature $F^{a}_{\mu\nu}$, which is exclusively consituted by light gauge fields $A^{(0)a}_\mu$. The presence of this covariant low--energy structure ensures that the ordinary four--dimensional Yang--Mills theory is contained in the KK Lagrangian ${\cal L}_{\rm YM}^{\rm 4\,dim}$, and suggests that the KK theory is governed by the ${\rm SU}_4(N)$ gauge group. The latter asseveration is indeed incomplete, for the gauge symmetry of the KK theory turns out to be richer, as we will see in the next section.

\section{Gauge symmetries of the Kaluza--Klein theory}
\label{gsKKt}
One of the main features of the four--dimensional KK Lagrangian ${\cal L}_{\rm YM}^{\rm 4\,dim}$ is, doubtless, gauge symmetry. Some authors~\cite{DDG2,MPR,U} attempted to find the appropriate gauge transformations governing the KK theory emerged from the five--dimensional Yang--Mills Lagrangian. However, they did not succeed, as they did not pay attention to the crucial role played by the gauge parameters. The first consistent discussion on the issue of gauge symmetry was done in Ref.~\cite{NT1}, where it was pointed out that the assumption of gauge parameters propagating in the extra dimension invariably requires the KK expansion of covariant objects instead of five--dimensional gauge fields. This approach to perform KK expansions has been recently extended~\cite{CNT} to the whole five--dimensional Standard Model.

The determination of the explicit form of the gauge variations governing the KK theory is crucial because they enter as an essential ingredient of the quantization procedure. They also carry valuable information about the nature of the fields of the theory, which can yield important phenomenological consequences~\cite{NT3}. The deduction of the gauge transformations of the KK theory can be achieved in, at least, three different manners~\cite{NT1}. One of them is analogous to the derivation of the KK curvatures that we carried out in the last section. Another one follows from the Dirac's method~\cite{Dirac}, which gives additional information of the gauge structure of the theory. Finally, they can be derived from the BRST transformations. Through the next two sections, we will discuss the obtainment of the gauge transformations of the KK theory by using the first and the second paths.

\subsection{Gauge transformations from Fourier analysis}
Consider the five--dimensional gauge variations
\begin{eqnarray}
\delta{\cal A}^a_\mu(x,y)&=&{\cal D}^{ab}_\mu\alpha^b(x,y),
\\ \nonumber \\
\delta{\cal A}^a_5(x,y)&=&{\cal D}^{ab}_5\alpha^b(x,y),
\end{eqnarray}
which gather all necessary information to determine the four--dimensional gauge--symmetry transformations. By substituting Eqs.(\ref{KKAmu}), (\ref{KKA5}) and (\ref{KKgp}), and then executing an analysis similar to the one used to find the KK curvatures, the gauge transformations of the KK theory are straightforwardly obtained. Their explicit expressions are~\cite{NT1}
\begin{eqnarray}
\label{zmgt}
\delta A^{(0)a}_\mu&=&D^{ab}_\mu\alpha^{(0)b}+gf^{abc}A^{(m)b}_\mu\alpha^{(m)c},
\\ \nonumber \\
\label{emgt}
\delta A^{(m)a}_\mu&=&gf^{abc}A^{(m)b}_\mu\alpha^{(0)c}+D^{(mn)ab}_\mu\alpha^{(n)b},
\\ \nonumber \\
\label{pgbgt}
\delta A^{(m)a}_5&=&gf^{abc}A^{(m)b}_5\alpha^{(0)c}+D^{(mn)ab}_5\alpha^{(n)b},
\end{eqnarray}
where
\begin{eqnarray}
D^{(mn)ab}_\mu&=&\delta^{mn}D^{ab}_\mu-gf^{abc}\Delta^{mrn}A^{(r)c}_\mu,
\label{KKDmu}
\\ \nonumber \\
D^{(mn)ab}_5&=&-\delta^{mn}\delta^{ab}\frac{m}{R}-gf^{abc}\Delta'^{mrn}A^{(r)c}_5.
\label{KKD5}
\end{eqnarray}
The object $D^{(mn)ab}_\mu$ is a sort of covariant derivative, which is not the case of $D^{(mn)ab}_5$, for it does not involve any derivative. A novel and quite remarkable trait of this KK theory is the possibility of splitting the gauge variations into two independent sets of transformations. The first one is obtained by taking the excited modes of the gauge parameters equal to zero, that is, $\alpha^{(m)a}=0$ for $m=1,2,\ldots$ This leads to the transformations
\begin{eqnarray}
\delta A^{(0)a}_\mu&=&D^{ab}_\mu\alpha^{(0)b},
\label{SGT1}
\\ 
\delta A^{(m)a}_\mu&=&gf^{abc}A^{(m)b}_\mu\alpha^{(0)c},
\label{SGT2}
\\
\delta A^{(m)a}_5&=&gf^{abc}A^{(m)b}_5\alpha^{(0)c},
\label{SGT3}
\end{eqnarray}
which are defined, exclusively, by the zero--mode gauge parameters. Notice that these variations transform the zero modes as ${\rm SU}_4(N)$ gauge fields, while the remaining fields are transformed as matter fields in the adjoint representation of the group. This behavior, which is consistent with the fact that the KK zero modes correspond to the ordinary four--dimensional Yang--Mills fields, has given~\cite{NT1} this transformations the name of {\it standard gauge transformations} (SGT). Now consider the case in which the zero modes of the gauge parameters are equal to zero ($\alpha^{(0)a}=0$). The resulting transformations, which receive~\cite{NT1} the name of {\it nonstandard gauge transformations} (NSGT), are given by
\begin{eqnarray}
\delta A^{(0)a}_\mu&=&gf^{abc}A^{(m)b}_\mu\alpha^{(m)c},
\label{NSGT1}
\\
\delta A^{(m)a}_\mu&=&D^{(mn)ab}_\mu\alpha^{(n)b},
\label{NSGT2}
\\
\delta A^{(m)a}_5&=&D^{(mn)ab}_5\alpha^{(n)b}.
\label{NSGT3}
\end{eqnarray}
The nature of the fields under this set of gauge transformations differs from that with respect to the SGT. Under these variations, the zero modes do not behave as gauge fields, but they transform in a way that resembles the transformation in the adjoint representation of the ${\rm SU}_4(N)$ group, but involving an infinite sum of KK modes. The form of the variations of the excited--modes fields $A^{(m)a}_\mu$ under the NSGT indicates that these fields are gauge fields under such transformations. The transformation rule of the scalar fields $A^{(m)a}_5$ involves the object $D^{(mn)ab}_5$, which does not contain derivatives, so that these fields do not behave as gauge fields. Consider the particular gauge transformation defined by taking the excited modes of the gauge parameteres to be~\cite{NT1} $\alpha^{(m)a}=(R/m)A^{(m)a}_5$. In this context, the scalars transform under the NSGT as $A^{(m)a}_5\to A'^{(m)a}_5=0$. This result is very important, as it explicitly proves that there exists a specific gauge such that the KK scalars $A^{(m)a}_5$ can be eliminated from the theory. These scalars are indeed pseudo--Goldstone bosons that are generated by the breaking of five--dimensional gauge symmetry that comes along with the compactification of the extra dimension. In fact, within this election of the gauge, note that 
\begin{equation}
\frac{1}{2}{\cal F}^{(m)a}_{\mu5}{\cal F}^{(m)a\mu}\hspace{0.001cm}_5\longrightarrow{\cal F}'^{(m)a}_{\mu5}{\cal F}'^{(m)a\mu}\hspace{0.001cm}_5=\frac{1}{2}\left( \frac{m}{R} \right)^2A^{(m)a}_\mu A^{(m)a\mu},
\end{equation}
so that such gauge fixing not only removes the scalars from the theory, but also defines the mass of the excited KK gauge bosons, as it occurs in the Higgs mechanism. This means that the degrees of freedom of the scalars have been eaten by the excited KK gauge fields, which in this manner have become massive. This is why we call this fields pseudo--Goldstone bosons. It is an interesting feature of the KK theory that the excited modes $A^{(m)a}_\mu$ are massive fields transforming as gauge fields (under the NSGT). 

\subsection{Gauge transformations from Dirac's method}
In this subsection we employ the Dirac's method, which allows us to derive the gauge trasformations shown in Eqs.(\ref{zmgt}), (\ref{emgt}) and (\ref{pgbgt}). These transformations can be then used to define the SGT and the NSGT, as we showed in the preceding section. In the mean, we derive the constraints of the theory, and once the first class constraints are known, we will define the Castellani gauge generator~\cite{C} and use it to determine the gauge transformations.

The generalized momenta of the theory are given by
\begin{eqnarray}
\pi^{(0)a}_\alpha&=&\frac{\partial{\cal L^{\rm 4 dim}_{\rm YM}}}{\partial \dot{A}^{(0)a\alpha}}={\cal F}^{(0)a}_{\alpha0},
\\ \nonumber \\
\pi^{(m)a}_\alpha&=&\frac{\partial{\cal L}_{\rm YM}^{\rm 4\,dim}}{\partial\dot{A}^{(m)a\alpha}}={\cal F}^{(m)a}_{\alpha0},
\\ \nonumber \\
\pi^{(m)a}_5&=&\frac{\partial{\cal L}_{\rm YM}^{\rm 4\,dim}}{\partial\dot{A}^{(m)a}_5}={\cal F}^{(m)a}_{05},
\end{eqnarray}
with the dotted fields representing velocities. By taking the index $\alpha$ as spatial, say, for instance, $\alpha=i$ ($i=1,2,3$), these momenta produce some defined velocities:
\begin{eqnarray}
\dot{A}^{(0)ai}&=&\pi^{(0)a}_i-D^{ab}_iA^{(0)b}_0-gf^{abc}A^{(m)b}_iA^{(m)c}_0,
\\ \nonumber \\
\dot{A}^{(m)ai}&=&\pi^{(m)a}_i-D^{(mn)ab}_iA^{(n)b}_0-gf^{abc}A^{(m)b}_iA^{(0)c}_0,
\\ \nonumber \\
\dot{A}^{(m)a}_5&=&\pi^{(m)a}_5-D^{(mn)ab}_5A^{(n)b}_0+gf^{abc}A^{(m)b}_5A^{(0)c}_0.
\end{eqnarray}
On the other hand, the case of a time index, $\alpha=0$, leads to the following primary constraints:
\begin{eqnarray}
\phi^{1\,(0)}_a&\equiv&\pi^{(0)a}_0\approx0,
\\ \nonumber \\
\phi^{1\,(m)}_a&\equiv&\pi^{(m)a}_0\approx0,
\end{eqnarray}
with the superscript ``$1$", in the left--hand side of these equations, standing for primary. The primary Hamiltonian is given by
\begin{equation}
H^{1}=\int d^3x\left( {\cal H}^1 +\lambda^{(0)a}\phi^{1\,(0)}_a+\lambda^{(m)a}\phi^{1\,(m)}_a \right),
\end{equation}
where $\lambda^{(0)a}$ and $\lambda^{(m)a}$ are Lagrange multipliers and the Hamiltonian density ${\cal H}^1$ is given by
\begin{eqnarray}
{\cal H}^1&=&\frac{1}{2}\pi^{(0)a}_i\pi^{(0)a}_i+\frac{1}{2}\pi^{(m)a}_i\pi^{(m)a}_i+\frac{1}{2}\pi^{(m)a}_5\pi^{(m)a}_5+A^{(0)a}_0D^{ab}_i\pi^{(0)b}_i
\\ \nonumber&&
+A^{(m)a}_0D^{(mn)ab}_i\pi^{(n)b}_i-A^{(n)b}_0D^{(mn)ab}_5\pi^{(m)a}_5-gf^{abc}\left( \pi^{(0)a}_iA^{(m)b}_iA^{(m)c}_0
\right.
\\ \nonumber&&
\left.
+\pi^{(m)a}_iA^{(m)b}_iA^{(0)c}_0-\pi^{(m)a}_5A^{(m)b}_5A^{(0)c}_0
 \right)+\frac{1}{4}\left( {\cal F}^{(0)a}_{ij}{\cal F}^{(0)a}_{ij}
\right.
\\ \nonumber&&
\left.
+{\cal F}^{(m)a}_{ij}{\cal F}^{(m)a}_{ij}+2{\cal F}^{(m)a}_{i5}{\cal F}^{(m)a}_{i5}
\right).
\end{eqnarray}
By imposing consistency conditions on the primary constraints,
\begin{eqnarray}
\dot{\phi}^{1\,(0)}_a&=&\left\{ \phi^{1\,(0)}_a,H^{1} \right\}\approx0,
\\ \nonumber \\
\dot{\phi}^{1\,(m)}_a&=&\left\{ \phi^{1\,(m)}_a,H^{1} \right\}\approx0,
\end{eqnarray}
there emerge some secondary constraints,
\begin{eqnarray}
\phi^{2\,(0)}_a&=&D^{ab}_i\pi^{(0)b}_i-gf^{abc}\left( \pi^{(m)b}_iA^{(m)c}_i+\pi^{(m)b}_5A^{(m)c}_5 \right)\approx0,
\\ \nonumber \\
\phi^{2\,(m)}_a&=&D^{(mn)ab}_i\pi^{(n)b}_i-D^{(mn)ab}_5\pi^{(n)b}_5-gf^{abc}\pi^{(0)b}_iA^{(m)c}_i\approx0,
\end{eqnarray}
but no velocities are determined. The Poisson parentheses between all the secondary constraints weakly vanish, as they are all proportional to secondary constraints. The explicit expressions are
\begin{eqnarray}
\left\{ \phi^{2\,(0)}_a(x),\phi^{2\,(0)}_b(x') \right\}&=&gf^{abc}\phi^{2\,(0)}_c(x)\delta(\vec{x}-\vec{x}'),
\\ \nonumber \\
\left\{ \phi^{2\,(0)}_a(x),\phi^{\,2(m)}_b(x') \right\}&=&gf^{abc}\phi^{2\,(m)}_c(x)\delta(\vec{x}-\vec{x}'),
\\ \nonumber \\
\left\{ \phi^{2\,(m)}_a(x),\phi^{2\,(n)}_b(x') \right\}&=&gf^{abc}\left( \delta^{mn}\phi^{2\,(0)}_c(x)+\Delta^{mrn}\phi^{2\,(r)}_c(x) \right)\delta(\vec{x}-\vec{x}').
\end{eqnarray}
All other Poisson brackets trivially vanish. As no new constraints arise and no more velocities are determined, the conclusion is that all of the constraints of the KK theory are first class. This result is consistent with the fact that the KK excited modes $A^{(m)a}_\mu$ are gauge fields because they are the only ones that generate constraints. The derivation of the constraints of the KK theory that we discussed through this subsection was carried out at the four--dimensional level. Another way to find these results is achieved~\cite{NT1} by considering the well known constraints of the four--dimensional Yang--Mills theory and generalizing them to five dimensions. After compactification, one can perform the necessary KK expansions of the five--dimensional constraints, then utilize Fourier analysis, and finally find~\cite{NT1} the same expressions that we just showed.

According to Dirac, the first--class constraints of a given gauge theory are the generators of gauge transformations. As we have already calculated the first--class constraints of the KK theory, we can find the gauge transformations by defining the Castellani's gauge generator~\cite{C} as
\begin{eqnarray}
{\cal G}&=&\int d^3z\left[ \left(
D^{ab}_0\alpha^{(0)b}+gf^{abc}A^{(m)b}_0\alpha^{(m)c} 
\right)\phi^{1\,(0)}_a-\alpha^{(0)a}\phi^{2\,(0)}_a
\right.
\\ \nonumber&&
\left.
+\left( gf^{abc}A^{(m)b}_0\alpha^{(0)c}+D_0^{(mn)ab}\alpha^{(n)b} \right)\phi^{1\,(m)}_a-\alpha^{(m)a}\phi^{2\,(m)}_a
\right],
\end{eqnarray}
where the gauge parameters $\alpha^{(0)a}$ and $\alpha^{(m)a}$ are only restricted to be soft. We then calculate the Poisson brackets of this generator with all the dynamic variables of the theory as
\begin{eqnarray}
\delta A^{(0)a}_\mu&=&\left\{ A^{(0)a}_\mu,{\cal G} \right\}=D^{ab}_\mu\alpha^{(0)b}+gf^{abc}A^{(m)b}_\mu\alpha^{(m)c},
\\ \nonumber \\
\delta A^{(m)a}_\mu&=&\left\{ A^{(m)a}_\mu,{\cal G} \right\}=gf^{abc}A^{(m)b}_\mu\alpha^{(0)c}+D^{(mn)ab}_\mu\alpha^{(n)b},
\\ \nonumber \\
\delta A^{(m)a}_5&=&\left\{ A^{(m)a}_5,{\cal G} \right\}=gf^{abc}A^{(m)b}_5\alpha^{(0)c}+D^{(mn)ab}_5\alpha^{(n)b},
\end{eqnarray}
which consistently coincide with Eqs. (\ref{zmgt}), (\ref{emgt}) and (\ref{pgbgt}). As before, taking $\alpha^{(n)a}=0$ ($\alpha^{(0)a}=0$) leads to the SGT (NSGT).

\subsection{Covariant objects and gauge invariance of the Kaluza--Klein theory}
At this point, one detail about gauge symmetry of the KK theory still remains pending. Some lines above, we discussed the KK expansion of five--dimensional covariant objects, which in this case were the KK curvatures ${\cal F}^a_{MN}$. This procedure produced a set of KK excitations of the curvatures, which we denoted by ${\cal F}^{(0)a}_{\mu\nu}$, ${\cal F}^{(m)a}_{\mu\nu}$ and ${\cal F}^{(m)a}_{\mu5}$. What we are going to discuss in this subsection is that such KK excited modes are~\cite{NT1} covariant objects under the two sets of four--dimensional gauge transformations, that is, under the SGT and the NSGT. This issue is very important, as it permits one to elegantly prove that the KK theory is gauge invariant under any of these KK gauge variations.

The gauge covariance of the KK excitations of the curvature is indeed encripted in the five--dimensional curvatures, which transform under the ${\rm SU}_5(N)$ gauge group as
\begin{eqnarray}
\delta{\cal F}^a_{\mu\nu}(x,y)&=&g_5f^{abc}{\cal F}^b_{\mu\nu}(x,y)\alpha^{c}(x,y),
\\ \nonumber \\
\delta{\cal F}^a_{\mu5}(x,y)&=&g_5f^{abc}{\cal F}^b_{\mu5}(x,y)\alpha^c(x,y).
\end{eqnarray}
Now, we insert the KK expansions of the curvatures and the gauge parameters into these expressions, and then we carry out a Fourier analysis as we did to obtain the KK curvatures in terms of the KK modes of the fields. This straightforwardly produces the following transformation laws:
\begin{eqnarray}
\delta{\cal F}^{(0)a}_{\mu\nu}&=&gf^{abc}\left( {\cal F}^{(0)b}_{\mu\nu}\alpha^{(0)c}+{\cal F}^{(m)b}_{\mu\nu}\alpha^{(m)c} \right),
\\ \nonumber \\
\delta{\cal F}^{(m)a}_{\mu\nu}&=&gf^{abc}\left( {\cal F}^{(m)b}_{\mu\nu}\alpha^{(0)c}+\left( \delta^{mn}{\cal F}^{(0)b}_{\mu\nu}+\Delta^{mrn}{\cal F}^{(r)b}_{\mu\nu} \right)\alpha^{(n)c} \right),
\\ \nonumber \\
\delta{\cal F}^{(m)a}_{\mu5}&=&gf^{abc}\left( {\cal F}^{(m)b}_{\mu5}\alpha^{(0)c}+\Delta'^{mrn}{\cal F}^{(r)b}_{\mu5}\alpha^{(n)c} \right).
\end{eqnarray}
By taking vanishing KK excited modes of the gauge parameters, that is, $\alpha^{(n)a}=0$, one can determine the way in which the KK excitations of the curvatures transform under the SGT. On the other hand, by imposing $\alpha^{(0)a}=0$ one obtains the NSGT laws of these KK curvatures. In both cases, the KK curvatures transform as covariant objects. The lesson to learn is that {\it KK expanding covariant objects in extra--dimensional gauge theories produces four--dimensional covariant objects}. The form of the gauge transformations of the KK curvatures allows one to prove in a rather simple way that the KK theory is gauge invariant:
$\delta{\cal L}_{\rm YM}^{\rm 4\,dim}=0$.
It is worth emphasizing that the four--dimensional KK Lagrangian ${\cal L}_{\rm YM}^{\rm 4\,dim}$ is invariant, separately, under each set of gauge transformations. This remarkable result is sublty related to the fact that the SGT are independent of the NSGT because they are defined by different sets of KK gauge parameters. The last assertion has another profound implication that affects quantization, which we will discuss in the next section.

\section{Quantization of the Kaluza--Klein theory}
\label{qKKt}
One of the main features of models involving UED is that they do not produce~\cite{UED1} tree--level corrections to low--energy observables. The absence of such tree--level contributions is a consequence of the so--called KK number conservation~\cite{UED1}, which occurs exclusively in UED contexts. If one assumes that there is no additional interaction that alters the momentum along the extra dimensions, the momentum, which is quantized by compactification of such directions, is stationary. At the level of the KK theory, this leads to a conservation of the KK number, which results in the absence of couplings involving only one KK excited mode and implies that the very first contributions to low--energy physics from the extra dimensions enter for the first time at the one--loop level. In this context, loop calculations containing UED effects play a prominent role. This occurs, for instance, in the case of gluon fusion $gg\to H$, which is generated for the first time at the one--loop level, so that it could receive~\cite{P} important contributions from extra--dimensional physics. The fact that the very first corrections to low--energy observables enter since the one--loop level in UED models also comes along with a relatively small lower bound on the compactification scale, which is estimated to be~\cite{UED1} $R^{-1}\gtrsim 300\,{\rm GeV}$. A lower bound as small as this could enhance extra--dimensional effects so that the International Linear Collider be sensitive to them, as it was shown in Ref.~\cite{FMNRT}. Such importance of radiative corrections, in this framework, renders the quantization of the KK excited modes a necessary requirement to perform calculations. As we discussed in the last section, the KK Lagrangian that emerged from the five--dimensional Yang--Mills theory is separately invariant under the SGT and the NSGT, which in turn are defined, respectively, by the zero modes and the excited modes of the gauge parameters. This means that these two sorts of gauge invariance are independent of each other, which is sublty related to the fact that the quantization of the KK theory can be divided~\cite{NT1} into two independent parts: the quantization of the KK zero modes; and the quantization of the KK excited modes. So one can quantize, for instance, the KK excited modes while leaving the KK zero modes as classical fields. We will take this route, as the calculation of extra--dimensional loop corrections of low--energy Green's functions only requires the quantization of the KK excited modes. Of course, once the KK excitations are quantized, one can do it with the zero modes. In what follows, we will discuss some general issues relative to the quantization of gauge systems within the BRST approach. At the same time, we will apply the main results to the KK theory.

\subsection{The master equation and its proper solution}
In general, the covariant quantization of gauge systems is achieved within the BRST formulation. The BRST symmetry emerges, classically, in the context of the Batalin--Vilkovisky formalism~\cite{BV}, which is also known as the field--antifield formalism~\cite{GPS}. Consider a gauge system described by an action $S_0[\phi]$, which is a functional of fields $\phi_i$. For simplicity, assume that the theory of interest is irreducible, so that there are no gauge transformations for gauge transformations. Ghost fields are commonly introduced in the quantization of gauge systems, as they are used to compensate for the effects of the gauge degrees of freedom~\cite{RF} in order to preserve unitarity. If the theory has $n$ gauge invariances, the formalism demands the inclusion, since the classical level, of $n$ ghost fields, $C^{k}$, that is, one ghost field per each gauge invariance of the theory. Indeed, such ghost fields coincide with the gauge parameters defining the gauge transformations, although their statistics is opposite to that of such parameters. In particular, the KK theory must comprise an infinite number of ghost fields, which is the number of gauge parameters defining gauge transformations. When GF and path integral quantization are considered, the minimal set, constituted by the fields $\phi_i$ and the ghost fields $C^k$, must be extended to a nonminimal set by introducing an antighost, $\bar{C}^k$, and an auxiliary field, $B^k$, for each ghost. Each couple $(\bar{C}^k,B^k)$ receives the name of {\it trivial pair}. We generically denote the fields constituting the non--minimal set by $\Phi^A$:
\begin{equation}
\Phi^A=\left\{ \phi_i,C^k,\bar{C}^k,B^k \right\}.
\end{equation}
Each of these fields is provided with an additive conserved charge, known as the {\it ghost number}. The ghost number is $0$ for matter, gauge and auxiliary fields, $+1$ for ghosts and $-1$ for antighosts. The number of dynamic variables of the theory is then further increased by introducing an antifield, $\Phi^*_A$, per each field $\Phi^A$. The statistics of a given antifield $\Phi^*_A$ is opposite to that of its corresponding field $\Phi^A$, and their ghost numbers are related by ${\rm gh}(\Phi^*_A)=-{\rm gh}(\Phi^A)-1$. The system so enlarged is described by an {\it extended action}, $S[\Phi,\Phi^*]$, which is a bosonic functional of the fields and antifields that has ghost number $0$.

The next step consists in endowing the configuration space with a symplectic structure that is known as the antibracket, which is defined in terms of left and right differentiations as
\begin{equation}
\left( F,G \right)=\frac{\partial_R F}{\partial \Phi^A}\frac{\partial_L G}{\partial \Phi_A^*}-\frac{\partial_R F}{\partial\Phi^*_A}\frac{\partial_LG}{\partial\Phi^A},
\end{equation}
where $F$ and $G$ are two functionals. The antifield $\Phi^*_A$ is canonically conjugate to the field $\Phi^A$, which means that they obey the fundamental antibrackets
\begin{eqnarray}
\left( \Phi^A,\Phi^*_B \right)&=&\delta^A_B,
\\ \nonumber \\
\left( \Phi^A,\Phi^B \right)=&0&=\left( \Phi^*_A,\Phi^*_B \right).
\end{eqnarray}
The extended action $S$ satisfies the master equation,
\begin{equation}
\left( S,S \right)=2\frac{\partial_RS}{\partial\Phi^A}\frac{\partial_LS}{\partial\Phi^*_A}=0,
\end{equation}
and is the generator of BRST transformations,
\begin{equation}
\delta_{\rm B}X=\left( X,S \right),
\label{BRSTt}
\end{equation}
where $X$ is a functional of fields and antifields. By virtue of the form of the BRST transformations, Eq.(\ref{BRSTt}), and the master equation, notice that the extended action $S$ is BRST invariant, that is, $\delta_{\rm B}S=0$. In general, a solution to the master equation has its own set of gauge invariances. When the number of such invariances equals the number of antifields, the solution receives the name of {\it proper solution}. An acceptable solution to the master equation must meet two indispensable requisites: (1) it must be a proper solution; and (2) it must make contact with the original action $S_0$ by satisfying the boundary condition
\begin{equation}
\left.S[\Phi,\Phi^*]\right|_{\Phi^*=0}=S_0[\phi].
\end{equation}
The proper solution to the master equation can be expanded in the antifields as
\begin{equation}
S[\Phi,\Phi^*]=S_0[\phi]+(\delta_{\rm B}\Phi^A)\Phi^*_A+\cdots,
\end{equation}
in which all the gauge--structure tensors characterizing the gauge system appear. In this sense, the proper solution $S$ is the generating functional of the gauge-structure tensors. The proper solution also generates the gauge algebra through the master equation. For this reason, the complete determination of a classical gauge system is achieved when the proper solution is established and the master equation is calculated, as these two steps generate the gauge structure tensors and the gauge algebra that they must satisfy. The gauge variations $\delta_{\rm B}\Phi^A$ are not known beforehand, but one can propose the most general proper solution to the master equation with gauge--structure tensors and then use it to obtain the variations.

Strictly speaking, the whole procedure discussed above should be followed to solve the classical KK system within the BRST framework. Nevertheless, there is a quite simpler path~\cite{NT1} that leads to the desired result. The idea is that the solution to the four--dimensional Yang--Mills theory is a known result that can be directly generalized to the five--dimensional case. After compactification, the corresponding five--dimensional proper solution can be manipulated through Fourier analysis to obtain the four--dimensional KK proper solution. The five--dimensional version of the proper solution to the master equation corresponding to the Yang--Mills theory is
\begin{eqnarray}
S=\int d^4x\int_0^{\pi R} dy\left( -\frac{1}{4}{\cal F}^a_{MN}{\cal F}^{aMN}+ {\cal A}^*_{aM}{\cal D}^{abM}C^b+\frac{1}{2}g_5f^{abc}C^*_cC^bC^a+\bar{C}^ *_aB_a \right),
\label{5DYMproper}
\end{eqnarray}
which clearly sataisfies the boundary condition $S|_{\Phi^*=0}=S_0$, that is, by eliminating the antifields $\Phi^*_A$ we recover the ordinary five--dimensional Yang--Mills action, defined by the ${\cal L}_{\rm YM}^{\rm 5\,dim}$ Lagrangian. The compactification of the extra dimension grants parity and periodicity properties to the dynamic variables and covariant objects, so that they can be expanded in KK towers. In the case of the first term of the proper solution, Eq.(\ref{5DYMproper}), Fourier analysis leads to the KK theory described by the ${\cal L}_{\rm YM}^{\rm 4\,dim}$ Lagrangian. Concerning the other terms, each antifield is assumed to have the same parity behavior than its corresponding field. In the case of the ghost fields, as they coincide with the gauge parameters, they have even parity under reflection of the extra--dimensional coordinate. Of course, the antifields of the ghost fields are assumed to be even, for their zero--modes are necessary to recover the proper solution to the ordinary four--dimensional Yang--Mills theory. This also serves as an argument to assume that the auxiliary fields $B_a$ are even. By KK--expanding covariant objects and integrating out the extra dimension, the following KK extended action is derived:
\begin{eqnarray}
\label{KKextS}
S&=&\int d^4x\Bigg[
{\cal L}_{\rm YM}^{\rm 4\,dim}+A^*_{(0)a\mu}D^{ab\mu}C^{(0)b}+\frac{1}{2}gf^{abc}C^*_{(0)c}C^{(0)b}C^{(0)a}
+\bar{C}^*_{(0)a}B^{(0)}_a
\\ \nonumber&&
+A^*_{(m)a\mu}D^{(mn)ab\mu}C^{(n)b}-A^*_{(m)a5}D^{(mn)ab}_5C^{(n)b}+\bar{C}^*_{(m)a}B^{(m)}_a+\frac{1}{2}gf^{abc}C^*_{(0)c}C^{(m)b}C^{(m)a}
\\ \nonumber&&
+\frac{1}{2}gf^{abc}C^*_{(m)c}\left( C^{(0)b}C^{(m)a}+C^{(0)a}C^{(m)b}+\Delta^{mrn}C^{(r)b}C^{(n)a} \right)
\Bigg].
\end{eqnarray}
Notice that this expression satisfies the proper--solution boundary condition. On the other hand, observe that the elimination of all terms involving KK excited modes leads to the proper solution of the four--dimensional Yang--Mills theory. Finally, it is worth remarking that the calculation of the variations $\delta_{\rm B}\Phi^A=(\Phi^A,S)$ allows one to consistently recover~\cite{NT1} the SGT, Eqs.(\ref{SGT1}), (\ref{SGT2}), (\ref{SGT3}), and the NSGT, Eqs.(\ref{NSGT1}), (\ref{NSGT2}), (\ref{NSGT3}). As the proper solution exhibited in Eq.(\ref{KKextS}) classically solves the KK theory, the next step is quantization, which we perform in the next subsection.

\subsection{The quantum Lagrangian}
In general, the quantization of a given gauge system calls for the fixation of the gauge. Gauge systems are enriched by the deep concept of gauge symmetry, whose origin is an overdescription~\cite{Dirac} in the sense that the number of degrees of freedom introduced to fully describe a given physical system surpasses the minimal number of degrees of freedom that is necessary to achieve the picture. Different gauges represent different mathematical configurations of the physical description that are related by gauge transformations. As the theory is invariant under such transformations, there is freedom to elect one particular gauge and perform reliable calculations. At the end of the day, physically acceptable results must not depend on the choice of the gauge. The fixation of the gauge is very important for quantization, as leaving the degeneration associated to gauge symmetry renders the path integral divergent, so that it can not be properly quantized. The BRST quantization incorporates the fixation of the gauge in a rather elegant manner, which we briefly describe below. 

Consider the following two facts: (1) the extended action, Eq.(\ref{KKextS}), possesses gauge symmetry, and hence cannot be directly quantized; (2) the antifields in such extended action do not represent physical degrees of freedom, so that they must be removed. Instead of just setting the antifields equal to zero, which would get the calculation back to the beginning, one can kill two birds with one stone by fixing the gauge through a device that, at the same time, removes the antifields. In general, the elimination of the antifields can be achieved by defining a GF fermion, $\Psi$, satisfying
\begin{equation}
\Phi^*_A=\frac{\partial\Psi}{\partial\Phi^A}.
\end{equation}
This fermion is a functional that depends on the fields and has ghost number $-1$. It is worth emphasizing that the requirement of properness of the solution to the master equation is vital for this GF procedure, as it ensures that such solution has the exact number of antifields to appropriately remove all the gauge degeneracy. On the other hand, notice that the presence of the trivial pairs is necessary because the only fields with ghost number $-1$ are the antighosts. Recall that the KK theory has two sorts of gauge symmetries, which are independent of each other. We wish to quantize only the KK excited modes, so that fixing the gauge with respect to the NSGT is enough. To do so, we introduce the fermionic functional $\Psi_{\rm NSGT}$, which will allow us to remove the antifields of the KK excited modes via the relation
\begin{equation}
\label{ff}
\Phi^*_{(m)A}=\frac{\partial\Psi_{\rm NSGT}}{\partial\Phi^{(m)A}}.
\end{equation}
The explicit form of $\Psi_{\rm NSGT}$ is
\begin{equation}
\Psi_{\rm NSGT}=\int d^4x\,\bar{C}^{(m)a}\left( f^{(m)a}+\frac{\xi}{2}B^{(m)a}+gf^{abc}\Delta^{mrn}\bar{C}^{(r)b}C^{(n)c} \right),
\end{equation}
wiht $\xi$ being the GF parameter and $f^{(m)a}$ representing bosonic GF functions, which we will conveniently define below. Using this expression and Eq.(\ref{ff}) yields the relations
\begin{eqnarray}
A^*_{(n)b\mu}&=&\frac{\partial f^{(m)a}}{\partial A^{(n)b\mu}}\bar{C}^{(m)a},
\\ \nonumber \\
A^*_{(n)b5}&=&\frac{\partial f^{(m)a}}{\partial A^{(n)b}_5}\bar{C}^{(m)a},
\\ \nonumber \\
C^*_{(m)a}&=&gf^{abc}\Delta^{mrn}\bar{C}^{(r)b}\bar{C}^{(n)c},
\\ \nonumber \\
\bar{C}^*_{(m)a}&=&f^{(m)a}+\frac{\xi}{2}B^{(m)}_a+2gf^{abc}\Delta^{mrn}\bar{C}^{(r)b}C^{(n)c},
\end{eqnarray}
which we employ to eliminate the antifields in the extended action $S$. The resulting expression is known as the gauge--fixed action, which we denote by $S_{\Psi_{\rm NSGT}}$. It si explicitly given by
\begin{eqnarray}
S_{\Psi_{\rm NSGT}}&=&\int d^4x\Bigg[ {\cal L}_{\rm YM}^{\rm 4\,dim}+A^*_{(0)\mu a}D^{ab\mu}C^{(0)b}
+\bar{C}^*_{(0)a}B^{(0)}_a
\\ \nonumber&&
+\frac{1}{2}gf^{abc}C^*_{(0)c}( C^{(0)b}C^{(0)a}
+C^{(m)a}C^{(m)b} )
\\ \nonumber&&
+\bar{C}^{(m)c}\frac{\partial f^{(m)c}}{\partial A^{(n)a\mu}}D^{(nr)ab\mu}C^{(r)b}
-\bar{C}^{(m)c}\frac{\partial f^{(m)c}}{\partial A^{(n)a}_5}D^{(nr)ab}_5C^{(r)b}
\\ \nonumber&&
+\frac{\xi}{2}B^{(m)}_aB^{(m)}_a
+B^{(m)}_a(f^{(m)a}+2gf^{abc}\Delta^{mrn}\bar{C}^{(r)b}C^{(n)c})
\\ \nonumber&&
+\frac{1}{2}g^2f^{abc}f^{cde}\Delta^{mpq}\bar{C}^{(p)d}\bar{C}^{(q)e}(C^{(0)b}C^{(m)a}+C^{(0)a}C^{(m)b}+\Delta^{mrn}C^{(r)b}C^{(n)a})
\Bigg].
\end{eqnarray}
It is worth empasizing that the gauge--fixed action $S_{\Phi_{\rm NSGT}}$ is not anymore invariant under the NSGT, although one can define suitable GF functions $f^{(m)a}$ that do not involve terms explicitly breaking invariance with repect to the SGT and, in this manner, maintain the symmetric behavior with respect to such transformations. The preservation of the SGT gauge symmetry will introduce important simplifications for subsequent calculations that we will perform in the next chapter. For such reason, we provide, below, a set of GF functions that will fulfill this requisite. Note that the auxiliary fields $B^{(m)}_a$ are still present in the gauge--fixed action, but they do not propagate. As they appear quadratically, we could integrate them out in order to remove them. However, such an integration is equivalent to directly using the equations of motion,
\begin{equation}
B^{(m)}_a=-\frac{1}{\xi}(f^{(m)a}+2gf^{abc}\Delta^{mrn}\bar{C}^{(r)b}C^{(n)c}),
\end{equation}
which we employ to eliminate these fields and obtain a quantum Lagrangian of the form
\begin{eqnarray}
\label{LKK}
{\cal L}_{\rm QKK}={\cal L}_{\rm YM}^{\rm 4\,dim}+{\cal L}_{\rm GF}+{\cal L}_{\rm FPG1}+{\cal L}_{\rm FPG2},
\end{eqnarray}
where the GF term, defined by the functions  $f^{(m)a}$, is given by
\begin{equation}
{\cal L}_{\rm GF}=-\frac{1}{2\xi}f^{(m)a}f^{(m)a},
\end{equation}
while the FPG part has been split into two parts as
\begin{eqnarray}
{\cal L}_{\rm FPG1}&=&\bar{C}^{(m)c}\left( \frac{\partial f^{(m)c}}{\partial A^{(n)a\mu}}D^{(nr)ab\mu}-\frac{\partial f^{(m)c}}{\partial A^{(n)a}_5}D^{(nr)ab}_5 \right)C^{(r)b}
-\frac{1}{\xi}gf^{abc}\Delta^{mrn}f^{(m)a}\bar{C}^{(r)b}C^{(n)c},
\\ \nonumber \\
{\cal L}_{\rm FPG2}&=&\frac{1}{2}gf^{abc}f^{cde}\Delta^{mpq}\bar{C}^{(p)d}\bar{C}^{(q)e}(C^{(0)b}C^{(m)a}
+C^{(0)a}C^{(m)b}+\Delta^{mrn}C^{(r)b}C^{(n)a}).
\end{eqnarray}
The second FPG part contains all the four-ghost fields terms, and is not important for the present work, so that we will forget about it from here on. On the other hand, notice that the GF term is not the only one that involves the GF functions, as they also permeate the ${\cal L}_{\rm FPG1}$ term. As we wish invariance of the theory under the SGT to be preserved, we introduce the following GF functions,
\begin{equation}
f^{(m)a}=D^{ab}_\mu A^{(m)b\mu}-\xi\frac{m}{R}A^{(m)a}_5,
\end{equation}
which transform covariantly under the SGT. This set of GF functions is indeed inspired by a proposal originaly given in Ref.~\cite{MRTT}, where a similar GF procedure was employed in the context of the 331 model~\cite{331}. By introducing these GF functions into the different terms of the ${\cal L}_{\rm QKK}$ Lagrangian, Eq.(\ref{LKK}), we obtain the following expressions for the GF and FPG parts:
\begin{eqnarray}
{\cal L}_{\rm GF}&=&-\frac{1}{2\xi}(D^{ab}_\mu A^{(m)b\mu})(D^{ac}_\nu A^{(m)c\nu})+m_mA^{(m)a}_5(D^{ab}_\mu A^{(m)b\mu})
-\frac{1}{2}\xi m_m^2A^{(m)a}_5A^{(m)a}_5,
\\ \nonumber \\
{\cal L}_{\rm FPG1}&=&\bar{C}^{(m)b}(D^{ba}_\mu D^{ac\mu})C^{(m)c}-\xi m_m^2\bar{C}^{(m)a}C^{(m)a}
-gf^{abc}\bigg[ \Delta^{mrn}\bar{C}^{(m)d}(D^{ad}_\mu A^{(r)c\mu})C^{(n)b}
\\ \nonumber&&
-\frac{1}{\xi}\Delta^{mrn}\bar{C}^{(r)c}(D^{ad}_\mu A^{(m)d\mu})C^{(n)b}
+\xi m_m\Delta'^{mrn}\bar{C}^{(m)a}A^{(r)c}_5C^{(n)b}-m_m\Delta^{mrn}\bar{C}^{(r)a}A^{(m)c}_5C^{(n)b}
\bigg],
\end{eqnarray}
where $m_m=m/R$ is the mass of the $m$--th KK excited mode. Notice also that both the pseudo--Goldstone bosons $A^{(m)a}_5$ and the ghost (antighost) fields $C^{(m)a}$ ($\bar{C}^{(m)a}$) have aquired unphysical masses $\sqrt{\xi}m_m$, that is, gauge--dependent masses.

Another interesting feature of this approach for the fixation of the gauge is that it leads to the elimination of some vertices involving pseudo--Goldstone bosons, which is convenient from the practical viewpoint, as loop calculations become easier. Such eliminations occur in the case of the unphysical bilinear and trilinear couplings $A^{(m)a}_\mu A^{(n)b}_5$ and $A^{(0)a}_\mu A^{(m)b}_\nu A^{(n)c}_5$ that are generated by the term $1/2{\cal F}^{(m)a}_{\mu5}{\cal F}^{(m)a\mu}\hspace{0.001cm}_5$. Such couplings are  cancelled by summing them with the GF part:
\begin{equation}
\frac{1}{2}{\cal F}^{(m)a}_{\mu5}{\cal F}^{(m)a\mu}\hspace{0.001cm}_5+{\cal L}_{\rm GF}=m_m\left[ A^{(m)a}_5(D^{ab}_\mu A^{(m)b\mu})+A^{(m)a\mu}(D^{ab}_\mu A^{(m)b}_5) \right]+\cdots=
m_m\partial_\mu(A^{(m)a}_5A^{(m)b\mu})+\cdots
\end{equation}

\section{One--loop renormalizability of extra--dimensional effects}
Extra--dimensional theories involve coupling constants with inverse--mass dimensions, which means that they are nonrenormalizable\footnote{Renormalizability of extra--dimensional theories has been discussed in Ref.~\cite{Gies} in a nonperturbative context.}. This happens, for instance, with the five--dimensional Yang--Mills theory, but notice that the KK theory that we derived by compactifying and integrating out the extra dimension is composed exclusively by terms whose canonical dimension is less or equal than four. If we had known the KK Lagrangian ${\cal L}_{\rm YM}^{\rm 4\,dim}$ without a prior knowledge of its extra--dimensional origin, at the first glance we could have na\"ively said, on the grounds of the Dyson's renormalizability criterion, that such theory is renormalizable. But then what happened with the nonrenormalizable nature of the five--dimensional Yang--Mills theory? The answer lies in one of the main consequences of compactification: the KK infinite sums. It is through the KK sums that the nonrenormalizability of the five--dimensional theory manifests, for, in general, such sums diverge. One way to get rid of such divergencies consists in truncating~\cite{CDH} the KK towers. This idea can be physically supported by thinking that the nonrenormalizable behavior of extra--dimensional theories indicates that a more fundamental description enters at an energy scale beyond the compactification scale,  but not too far from it, which in turn implies that one can take into account only a finite number of terms of the series. The five--dimensional version of the SM is an interesting particular case, for each one--loop contribution to low--energy physics involves only one~\cite{UED1,NT1} KK infinite sum, which produces Riemman--$\zeta$ functions. Such series are convergent, so that the one--loop corrections to light Green's functions are~\cite{NT1,NT2} renormalizable. The possibility of renormalizing such contributions is important because they incarnate the lowest--order effects of the extra dimension on light physics, which means that some of the largest corrections to low--energy physics can be calculated without amiguities. It is worth appreciating the great relevance of this result. In this section, we provide a proof~\cite{NT1} of the renormalizability of the one--loop effects of the extra dimension on light Green's functions. We will show that the divergencies emerged from the extra--dimensional effects can be absorbed by the parameters of the low--energy theory.

As a first step, we quantize the ordinary four--dimensional Yang--Mills theory, for which we must fix the gauge with respect to the zero--mode gauge fields $A^{(0)a}_\mu$. The proof of one--loop renormalizability of the standard Yang--Mills theory is more simply achieved by using the Background Field Method~\cite{BFM} (BFM), which ensures preservation of gauge invariance with respect to the SGT. Within such framework, we divide the zero--mode gauge fields $A^{(0)a}_\mu$ into two parts, 
\begin{equation}
A^{(0)a}_\mu\to A^{(0)a}_\mu+{\cal Q}^{(0)a}_\mu,
\end{equation}
where the $A^{(0)a}_\mu$ in the right--hand side is a classical background field and ${\cal Q}^{(0)a}_\mu$ is a fluctuating quantum field. In this context, the classical background field $A^{(0)a}_\mu$ is assumed to be a classical field configuration, while ${\cal Q}^{(0)a}_\mu$ is a functional integration variable. By using this splitting, the curvature is replaced by
\begin{equation}
F^a_{\mu\nu}\to F^a_{\mu\nu}+D^{ab}_\mu{\cal Q}^{(0)b}_\nu-D^{ab}_\nu{\cal Q}^{(0)b}_\mu+gf^{abc}{\cal Q}^{(0)b}_\mu{\cal Q}^{(0)c}_\nu.
\end{equation}
Now consider the GF condition
\begin{equation}
f^{(0)a}=D^{ab}_\mu{\cal Q}^{(0)b\mu},
\end{equation}
which is SGT--covariant. The gauge--fixed Lagrangian for the standard Yang--Mills theory is then given by
\begin{eqnarray}
{\cal L}^{(0)}_{\rm YM}&=&-\frac{1}{4}\left( F^a_{\mu\nu}+D^{ab}_\mu{\cal Q}^{(0)b}_\nu-D^{ab}_\nu{\cal Q}^{(0)b}_\mu+gf^{abc}{\cal Q}^{(0)b}_\mu{\cal Q}^{(0)c}_\nu \right)^2
\\ \nonumber&&
-\frac{1}{2\xi}(D^{ab}_\mu{\cal Q}^{(0)b\mu})^2+\bar{C}^{(0)a}(D^{ab}_\mu D^{bd\mu}+gf^{bcd}D^{ab\mu}{\cal Q}^{(0)c}_\mu)C^{(0)d}.
\end{eqnarray}
This Lagrangian is invariant under the SGT, with the ghost and fluctuating quantum fields transforming in the adjount representation of the ${\rm SU}_4(N)$ gauge group.

The quantization of the low--energy theory, performed through the above lines, has completed the quantization of the whole KK theory. From the complete KK quantum Lagrangian, we extract those parts that contribute at the one--loop level to light Green's functions, and find that the corresponding terms are
\begin{equation}
{\cal L}_{\rm 1-loop}^{\xi}={\cal L}_{\rm 1-loop}^{(0)\xi}+\sum_{m=1}^\infty{\cal L}_{\rm 1-loop}^{(m)\xi},
\end{equation}
with
\begin{eqnarray}
{\cal L}_{\rm 1-loop}^{(0)\xi}&=&-\frac{1}{2}\left( \frac{1}{2}(D^{ab}_\mu{\cal Q}^{(0)b}_\nu-D^{ab}_\nu{\cal Q}^{(0)b}_\mu)^2+gf^{abc}F^{a\mu\nu}{\cal Q}^{(0)b}_\mu{\cal Q}^{(0)c}_\nu
\right.
\\ \nonumber&&
\left.
+\frac{1}{\xi}(D^{ab}_\mu Q^{(0)b\mu})^2\right)+\bar{C}^{(0)b}(D^{ba}_\mu D^{ac\mu})C^{(0)c},
\\ \nonumber \\
\label{1loopxi}
{\cal L}_{\rm 1-loop}^{(m)\xi}&=&-\frac{1}{2}\left( \frac{1}{2}(D^{ab}_\mu A^{(m)b}_\nu-D^{ab}_\nu A^{(m)b}_\mu)^2+gf^{abc}F^{a\mu\nu}A^{(m)b}_\mu A^{(m)c}_\nu \right.
\\ \nonumber &&
\left. +\frac{1}{\xi}(D^{ab}_\mu A^{(m)b\mu})^2-m_m^2A^{(m)a}_\mu A^{(m)a\mu} \right)
\\ \nonumber&&
+\bar{C}^{(m)b}(D^{ba}_\mu D^{ac\mu}-\xi m_m^2)C^{(m)c}+\frac{1}{2}\left((D^{ab}_\mu A^{(m)b}_5)^2-m_m^2A^{(m)a}_5A^{(m)a}_5 \right).
\end{eqnarray}
It is worth remarking that the ${\cal L}_{\rm 1-loop}^{(0)\xi}$ and ${\cal L}_{\rm 1-loop}^{(m)\xi}$ are quite similar. Notice also that the couplings of the ${\cal L}_{\rm 1-loop}^{(m)\xi}$ Lagrangian are all of renormalizable type and that they are all the ones allowed by gauge invariance. This in turn implies that the divergences generated the KK modes $A^{(m)a}_\mu$ must have the same structure of those produced by the zero--mode quantum fluctuations ${\cal Q}^{(0)a}_\mu$. It occurs that the gauge invariance left by the background field gauge ensures that the ultra--violet divergencies (UVd) associated with the standard Yang--Mills theory  have the form
\begin{equation}
{\cal L}_{\rm UVd}^{(0)}=-\frac{1}{4}L^{(0)}F^a_{\mu\nu}F^{a\mu\nu},
\end{equation}
where dimensional analysis indicates that $L^{(0)}$ is logarithmically divergent. As just commented, this is the structure of the divergencies that the KK excited modes are expected to produce, so that the total one--loop KK divergencies, which include those associated to both the zero and the excited modes, are expressed as
\begin{equation}
{\cal L}_{\rm UVd}=-\frac{1}{4}\sum_{m=0}^\infty L^{(m)}F^a_{\mu\nu}F^{a\mu\nu}\equiv-\frac{1}{4}LF^a_{\mu\nu}F^{a\mu\nu},
\end{equation}
with $L^{(0)}=L^{(1)}=\cdots =L^{(m)}=\cdots$. The divergencies generated by the KK excited modes can be then absorbed by the parameters of the low--energy theory by defining the renormalized zero--mode fields, $A^{{\rm R}(0)a}_\mu$, and coupling constant, $g_{\rm R}$, as
\begin{eqnarray}
A^{{\rm R}(0)a}_\mu&=&(1+L)^{\frac{1}{2}}\,A^{(0)a}_\mu,
\\ \nonumber \\
g_{\rm R}&=&(1+L)^{-\frac{1}{2}}g,
\end{eqnarray}
which produce the renormalized curvature
\begin{equation}
F_{\mu\nu}^{{\rm R}a}=\partial_\mu A^{{\rm R}(0)a}_\mu-\partial_\nu A^{{\rm R}(0)a}_\mu+g_{\rm R}f^{abc}A^{{\rm R}(0)b}_\mu A^{{\rm R}(0)c}_\nu.
\end{equation}
Two main ingredients of this proof of renormalization are the gauge invariance left by the background field gauge and the covariant GF procedure introduced for the KK excited modes. It is a well--known fact, from radiative corrections, that logarithmically divergent integrals can introduce effects that are proportional to the logarithm of the mass of the particle circulating in the loop. In the case of this KK theory, such logarithmic divergence should have the form ${\rm log}(mR^{-1}/\mu)$, where $\mu$ is a mass scale like the one inroduced by dimensional regularization and $m$ is a KK mode number. The nondecoupling behavior of the terms involving such logarithmic divergence are irrelevant as they are unobservable, for they can be absorbed by renormalization. In the next chapter we will prove renormalizability of the KK one--loop contributions by integrating out~\cite{NT2} the KK excited modes.

\chapter{Integration of Kaluza--Klein modes}
\label{KKint}
The integration of the KK excited modes originated in the five--dimensional version of the Yang--Mills theory and the posterior derivation of an effective Lagrangian including the first KK excited--modes one--loop effects has alluring features that deserve some comments. As we discussed above, the KK excited modes are gauge fields, so that the quantization of them required a GF procedure, which inserted a GF parameter, $\xi$, into the theory. The possibility that extra--dimensional physics produces an effective Lagrangian expansion that depends on the GF procedure comes into the game. This interesting issue had never been discussed in the literature, as the integration of heavy gauge fields to obtain an effective Lagrangian is a novel calculation. From the phenomenological viewpoint, the integration of KK modes is attractive because, as we have already discussed, the contributions of this sort of heavy physics on light Green's functions appear for the first time at the one--loop level. The effective Lagrangian resulting from the integration of the KK excited modes shall comprehend such one--loop effects, so that from this expansion one can obtain corrections to low--energy physics from tree--level diagrams. As we proved above, the one--loop effects from extra--dimensional physics produced by this model are renormalizable. This result can be also achieved by integrating out the KK excited modes, as we shall show below. The nonrenormalizable behavior of extra-dimensional models is physically important, as it means that there is doubtless a more fundamental theory, perhaps string theory, whose effects shall manifest at a higher energy scale, $M_{\rm S}$. The significance of this more fundamental description can be in principle parametrized through a five--dimensional effective Lagrangian constituted by the five--dimensional Yang--Mills theory and a sum of terms involving operators of canonical dimension higher than five. As the theory is nonrenormalizable, there is no criterion that restricts the number of such operators that one can include. The general form of the effective Lagrangian obtained by integrating out the KK excited modes allows one to perform~\cite{NT2} in a rather simple way a comparison of the extra--dimensional effects, characterized by the compactification scale $R^{-1}$, with those from the fundamental description of nature that lies beyond the compactification scale, and whose distinctive energy scale is $M_{\rm S}$. The integration of heavy KK modes is not, in general, an easy task, for even the simplest cases represent technical challenges. This is the case, for instance, of the Euler--Heisenberg Lagrangian~\cite{EH}, which is obtained by assuming that the electron field of the Lagrangian describing quantum electrodynamics is heavy, which allows one to integrate it~\cite{DMNP} out. The sole replacement~\cite{tA} of the abelian theory by the Yang--Mills theory introduces significative difficulties, rendering the calculation of the low--energy effective Lagrangian rather intricate. Instead of performing such a baroque calculation, as one should do by following the ordinary path, people has developed~\cite{hfim,BS} methods to systematically obtain effective Lagrangians in simplier ways. In particular, there is one method propounded~\cite{BS} by M. Bilenki and A. Santamaria some years ago. In that time, these authors considered an economic SM extension that incorporated a heavy scalar singlet coupled to the leptonic doublet. They supposed that the scalar was heavy, then integrated it out, and finally calculated an effective Lagrangian. In the present paper, we adjust their elegant method to the case of heavy gauge fields and obtain the low--energy effective description containting up to canonical--dimension--six nonrenormalizable operators built of light Yang--Mills covariant objects and being governed by the SGT, which represent the low--energy symmetry. As we will show, the proof will be very profitable in the sense that some interesting results will emerge along the process. It is worth commenting that the integration of heavy KK modes and the consequent obtainment of an effective Lagrangian expansion is an issue rarely considered in the literature~\cite{RW}, but it is however interesting because it provides a way to calculate the lowest--order extra--dimensional effects on light physics by dealing just with tree--level vertices instead of performing one--loop calculations. This issue enlarges the value of this work.

In the last chapter, we quantized the low--energy theory by means of the BFM. In what follows we will disregard such result and look only to the gauge--fixed Lagrangian ${\cal L}_{\rm QKK}$, Eq.(\ref{LKK}), where the only fields that have been subjected to quantization are the KK excited modes. By virtue of the goals pursued in this paper, we conveneintly rewrite this Lagrangian as
\begin{equation}
{\cal L}_{\rm QKK}={\cal L}_{\rm YM}+{\cal L}_{\rm 1-loop}^{{\rm KK},\xi}+{\cal L}_{\rm heavy},
\label{Lefflmh}
\end{equation}
where the ${\cal L}_{\rm YM}$ term is the ordinary four--dimensional Yang--Mills Lagrangian, which is given by
\begin{equation}
{\cal L}_{\rm YM}=-\frac{1}{4}F^a_{\mu\nu}F^{a\mu\nu}.
\end{equation}
The Lagrangian ${\cal L}_{\rm 1-loop}^{{\rm KK},\xi}$, which is a link between light physics and high energy extra--dimensional physics, can be written as
\begin{equation}
{\cal L}_{\rm 1-loop}^{{\rm KK},\xi}=\sum_{m=1}^\infty{\cal L}_{\rm 1-loop}^{(m)\xi},
\end{equation}
where the terms ${\cal L}_{\rm 1-loop}^{(m)\xi}$ were given in Eq.(\ref{1loopxi}). It comprises all one--loop effects of extra--dimensional physics on the light Green's functions, and hence on low--energy observables. As we discussed before, the one--loop corrections of extra--dimensional physics manifests for the first time at the one--loop level, so that the ${\cal L}_{\rm 1-loop}^{\rm KK,\xi}$ part contains all the first contributions of this UED model to low--energy physics. This term is particulartly relevant for the present work because it is the one from which the KK excited modes shall be integrated out to obtain the low--energy effective Lagrangian expansion. The label $\xi$ of the ${\cal L}^{\rm KK,\xi}_{\rm 1-loop}$ Lagrangian corresponds to the GF parameter, which is embedded in different parts of this Lagrangian and denotes gauge dependence. The explicit expression of the one--loop Lagrangian ${\cal L}_{\rm 1-loop}^{\rm KK,\xi}$, which shall be useful in the next section, is
\begin{eqnarray}
{\cal L}_{\rm 1-loop}^{{\rm KK},\xi}&=&\frac{1}{2}g_{\mu\nu}A^{(m)b\mu}D^{ba}_\alpha D^{ad\alpha}A^{(m)d\nu}+gf^{bad}A^{(m)b\mu}F^a_{\mu\nu}A^{(m)d\nu}
\\ \nonumber&&
-\frac{1}{2}\left( 1-\frac{1}{\xi} \right)A^{(m)b\mu}D^{ba}_\mu D^{ad}_\nu A^{(m)d\nu}
+\frac{1}{2}m_m^2g_{\mu\nu}A^{(m)a\mu}A^{(m)a\nu}
\\ \nonumber &&
-\frac{1}{2}A^{(m)b}_5D^{ba}_\alpha D^{ad\alpha}A^{(m)d}_5-\frac{1}{2}\xi m_m^2A^{(m)a}_5A^{(m)a}_5
\\ \nonumber &&
-\bar{C}^{(m)b}D^{ba}_\alpha D^{ad\alpha}C^{(m)d}
-\xi m_m^2\bar{C}^{(m)a}C^{(m)a},
\end{eqnarray}
Notice that the GF parameter $\xi$ has permeated all the one--loop Lagrangian ${\cal L}_{\rm 1-loop}^{{\rm KK},\xi}$. This observation is very important, as this spreading is the seed of the gauge dependence that is expected to appear in the final result. The last term of Eq.(\ref{Lefflmh}), which we denoted by ${\cal L}_{\rm heavy}$, produces contributions that enter into light Green's functions at the two--loop level and higher orders. This heavier--physics Lagrangian is irrelevant for the KK--modes integration that we are going to perform, so that we will ignore it from here on and only conserve the first two terms of the right--hand side of Eq.(\ref{Lefflmh}).

\section{The effective action}
The starting point is the effective action $S^{\xi}$, which we define as
\begin{eqnarray}
{\rm exp}\left\{ iS^\xi \right\}&=&\int DA^{(n)}_\mu DA^{(n)}_5D\bar{C}^{(n)}DC^{(n)}\,{\rm exp}\{ iS_{\rm QKK} \}
\\ \nonumber &
=&\int DA^{(n)}_\mu DA^{(n)}_5D\bar{C}^{(n)}DC^{(n)}\,{\rm exp}\left\{ i\int d^4x\,{\cal L}_{\rm QKK} \right\},
\end{eqnarray}
where a functional integration over all the KK excited modes has been indicated. This ensures that the final result shall depend only on the KK zero modes. As the full quantum Lagrangian ${\cal L}_{\rm QKK}$ is gauge invariant under the SGT, the integration of the KK excitations shall preserve this invariance, so that the nonrenormalizable terms of the final effective Lagrangian expansion shall be SGT--invariant. After integrating out the KK excited modes by solving some gaussian integrals, the following low--energy effective action arises:
\begin{eqnarray}
\label{Seff1/2}
S^\xi&=&S_{\rm YM}+\frac{i}{2}\sum_{m=1}^\infty{\rm Tr}\;{\rm log}\left[ g_{\mu\nu}(D^2+m_m^2)-\left( 1-\frac{1}{\xi} \right)D_\mu D_\nu-4igF_{\mu\nu} \right]
\\ \nonumber &&
+\frac{i}{2}\sum_{m=1}^\infty{\rm Tr}\;{\rm log}\left[ \frac{1}{2}\left(-D^2-\xi m_m^2\right) \right]-i\sum_{m=1}^\infty{\rm Tr}\;{\rm log}\left[ -D^2-\xi m_m^2 \right],
\end{eqnarray}
with $F_{\mu\nu}= F^a_{\mu\nu}T^a$, where $T^a$ represents the ${\rm SU}_4(N)$ gauge--group generators. Observe that we have denoted $D^2\equiv D_\mu D^\mu$. The symbol ``Tr" has been utilized to represent a trace affecting both the internal and the external degrees of freedom. External degrees of freedom correspond to the the four--dimensional spacetime points, which are labeled by continuous indices. On the other hand, the internal degrees of freedom are defined by the gauge and the four--dimensional Lorentz groups. The first term of Eq.(\ref{Seff1/2}) is the action for the four--dimensional Yang--Mills theory. All other parts of the effective action comprise one--loop level corrections of the KK excited modes to light Green's functions. The second term is the contribution of the KK excited modes $A^{(n)a}_\mu$, and is the only one that carries Lorentz indices. The third term comes from the integration of the pseudo--Goldstone bosons $A^{(n)a}_5$, while the fourth term was generated by the KK excited modes of the ghost fields. In obtaining this result, the GF procedure that we introduced for the KK excited modes was convenient, as it eliminated terms that mix pseudo--Goldstone bosons $A^{(m)a}_5$ with gauge KK excited modes $A^{(m)a}_\mu$, and the result was a great simplification of the integration of the heavy fields. There is another noteworthy attribute of Eq.(\ref{Seff1/2}) that is connected to the GF procedure for the KK excitations. By looking at the second term of such an expression, notice that the factor $1/2$ in the argument of the logarithm is inoffensive, as it contribures only to vacuum energy. With this in mind, it is evident that the arguments of the logarithms in the second and third terms are equal to each other. As it was mentioned few lines above, the third and fourth terms come, respectively, from the integration of the KK scalar excitations, $A^{(m)a}_5$, and the KK excitations of the ghost fields, $\bar{C}^{(m)a}$ and $C^{(m)a}$, so that this intermediate step shows that the following relation holds: {\it in the SGT--covariant gauge, the one-loop contributions to low--energy physics generated by the KK pseudo--Goldstone bosons are minus twice those produced by the KK excited ghost fields}. It is worth emphasizing that, as the last sentence asseverates, this result is an implication of the GF procedure that we introduced. In fact, this property has manifested in other contexts~\cite{MRTT,RTT} than extra dimensions in which analogous GF approaches, leaving some gauge invariance, have been employed. Concerning extra dimensions, this result has been exploited in phenomenological one--loop level calculations~\cite{FMNRT}, and it has provided technical advantages by leading to valuable simplifications. A final remark on this intermediate result is that this expression gets the simpler it can when taking the FtH gauge, that is $\xi=1$, as such a choice eliminates the cross-derivative term $D_\mu D_\nu$. We will take advantage of this fact in the next section, in which we will consider that gauge and then calculate the effective Lagrangian.

\section{The Feynman--`t Hooft gauge}
As we just mentioned, the FtH gauge simplifies the form of the effective action $S^\xi$ as no other election of the GF parameter does. Gauge systems, which supply elegant and fundamental descriptions of nature, are captivating, as they involve the enthralling concepts of gauge invariance and gauge independence. The former is indeed an essential piece in the construction of this sort of physical models. This symmetry relates different mathematical configurations that must be physically equivalent, so that the choice of any gauge must lead to the same physical results. The quantization of gauge systems requires the election of a particular gauge, for leaving this invariance renders the path integral divergent. The fixation of the gauge can be parametrized by means of a GF parameter which is expected to vanish when constructing $S$--matrix elements because all gauges are physically equivalent, although Green's functions could have such parameter incrustated in a nontrivial way. However, the correct combination of gauge dependent Green's functions to build a physical quantity is expected to eliminate this parameter. In general, the FtH gauge seems to be especial among all other possibilities, as it is physically interesting and convenient from a practical perspective. The physical importance of this gauge can be appreciated, for instance, by looking at the subtle relation between the BFM and the Pinch Technique~\cite{PT} (PT). As we discussed in the last chapter, the BFM propounds a split of the gauge fields, ${\cal G}^a_\mu$, of a given theory into a classical background, $G^a_\mu$, and a quantum fluctuation, $Q^a_\mu$, as ${\cal G}^a_{\mu}\to G^a_\mu+Q^a_\mu$. One then quantizes the $Q^a_\mu$ field, but leaves the $G^a_\mu$ field classical. Such a procedure allows one to conserve certain ``amount" of gauge invariance, which in a practical sense is desirable by virtue of the latent simplifications provided by symmetries. The quantization of the quantum fluctuation $Q^a_\mu$ requires to fix the gauge with respect to such gauge fields, which introduces a GF parameter, $\xi_Q$. The combination of all these issues yields Green's functions satisfying simple QED--like ward indentities but carrying gauge dependence through the presence of the GF parameter $\xi_Q$. On the other hand, the PT is a diagrammatic method that pursues the construction of a quantum action that leads to both gauge invariant and gauge independent Green's functions. This method consists in constructing well--behaved Green's functions of a given number of points by combining some individual contributions from Green's functions with equal and higher number of points, whose Feynman rules are derived from a conventional effective action or even from a nonconventional scheme. It was first established~\cite{BFMPT1} at the one--loop level that the Green's functions calculated by using the BFM surprisingly coincide with those obtained through the PT when working in the $\xi_{Q}=1$ gauge. This interesting finding was then proved~\cite{BFMPT2} to be valid at the two--loop level, and eventually it was found to be fulfilled~\cite{BFMPT3} at all orders of perturbation theory. Such an outstanding result suggests a subtle and fundamental connection among the BFM and the PT. So far, this link remains unexplained, but it is worth emphasizing  the physically relevant role played by the FtH gauge. Finally, we want to comment that this gauge does not produce unphysical thresholds and can be used to produce well--behaved results, even in the case of gauge--dependent Green's functions, as it was shown~\cite{FMNRT} recently for a KK theory in the context of UED.

Consider the effective action $S^\xi$, Eq.(\ref{Seff1/2}), in the FtH gauge ($\xi=1$). The pure--gauge trace is greatly simplified by  the elimination of the cross--derivative term provided  by this election, and the resulting effective action reads
\begin{equation}
S^1=S_{\rm YM}+\frac{i}{2}\sum_{m=1}^\infty{\rm Tr}\;{\rm log}\left[ g_{\mu\nu}(D^2+m_m^2)-4igF_{\mu\nu} \right]-\frac{i}{2}\sum_{m=1}^\infty{\rm Tr}\;{\rm log}\left[ -D^2-m_m^2 \right].
\end{equation}
The third term, in the right--hand side, looks like the trace corresponding to the integration of a scalar field, and it has been already solved in Ref.~\cite{BS}. In order to solve the traces and derive the effective Lagrangian, we will employ the dimensional regularization scheme, so that we will work in $d$ dimensions from here on. In $d$ dimensions, the first term of the argument of the logarithm in the pure--gauge trace satisfies
\begin{equation}
{\rm Tr}\;{\rm log}\left[ g_{\mu\nu}(D^2+m_m^2) \right]=d\,{\rm Tr}\;{\rm log}\left[ D^2+m_m^2 \right]
\end{equation}
which means that Lorentz indices introduce a global factor $d$ with respect to the sole scalar contributions. This expression should read reasonable by taking into account that a term like the one in the right--hand side should come from integrating out gauge fields, and that such objects are arranged as vectors with $d$ components, with each one of them contributing as a scalar. In Appendix \ref{app1} we calculate a more general pure--gauge trace, for which we find the following expansion:
\begin{eqnarray}
\label{pgtexp}
i\,{\rm Tr}\;{\rm log}\left[ g_{\mu\nu}(D^2+M^2)+U_{\mu\nu}(x) \right]&=&\int d^dx\Bigg[ \frac{1}{(4\pi)^2}M^2\left( \Delta_\epsilon +{\rm log}\left( \frac{\mu^2}{M^2} \right)+1 \right){\rm tr}\{ U^\mu\hspace{0.001cm}_\mu \}
\\ \nonumber&&
+\frac{1}{(4\pi)^2}\frac{1}{2}\left( \Delta_\epsilon+{\rm log}\left( \frac{\mu^2}{M^2} \right) \right){\rm tr}\{ U_{\mu\nu}U^{\mu\nu} \}
\\ \nonumber&&
-\frac{g^2}{(4\pi)^2}\frac{1}{3}\left( \Delta_\epsilon+{\rm log}\left( \frac{\mu^2}{M^2} \right)-\frac{1}{2} \right){\rm tr}\{ F_{\mu\nu}F^{\mu\nu} \}\Bigg]
\\ \nonumber&&
+\int d^4x\Bigg[-\frac{1}{(4\pi)^2}\frac{1}{6}\frac{1}{M^2}{\rm tr}\{ U_{\mu\nu}U^{\nu\sigma}U_\sigma\hspace{0.001cm}^\mu \}
\\ \nonumber&&
+\frac{1}{(4\pi)^2}\frac{1}{3}\frac{1}{M^2}{\rm tr}\{ D_\mu U^{\mu\nu}D^\sigma U_{\sigma\nu} \}
\\ \nonumber&&
-\frac{g^2}{(4\pi)^2}\frac{1}{3}\frac{1}{M^2}{\rm tr}\{ F_{\mu\nu}U^{\nu\sigma}F_\sigma\hspace{0.001cm}^\mu \}
\\ \nonumber&&
-\frac{g^2}{(4\pi)^2}\frac{1}{15}\frac{1}{M^2}{\rm tr}\{ D_\mu F^{\mu\nu}D^\sigma F_{\sigma\nu} \}
\\ \nonumber&&
-\frac{ig^3}{(4\pi)^2}\frac{2}{45}\frac{1}{M^2}{\rm tr}\{ F_{\mu\nu}F^{\nu\sigma}F_\sigma\hspace{0.001cm}^\mu \} \Bigg]+{\cal O}(1/M^4),
\end{eqnarray}
where the factor $\Delta_\epsilon$, which contains all divergencies for $d\to4$, is defined as
\begin{eqnarray}
\Delta_\epsilon=\frac{1}{\epsilon}-\gamma_{\rm E}+{\rm log}(4\pi),&&\epsilon=\frac{4-d}{2}.
\end{eqnarray}
The object $U_{\mu\nu}(x)$ is an arbitrary matrix--valued function of the spacetime coordinates and $\mu$ is a factor with dimension of mass that is introduced as part of dimensional regularization to correct units. To obtain this formula, all traces over spacetime indices and Lorentz indices as well were performed, so that the remaining trace, which was indicated by ``tr", concerns only gauge--group generators. It is worth pointing out that this expansion is rather general, as it can be employed to calculate effective Lagrangians from pure--gauge traces possessing the same structure, even in frameworks different than extra dimensions. The authors of Ref.~\cite{BS} gave the following expression for the scalar trace:
\begin{eqnarray}
i{\rm Tr}\;{\rm log}\left[ -(D^2+M^2) \right]&=&\int d^dx\Bigg[-\frac{g^2}{(4\pi)^2}\frac{1}{12}\left( \Delta_\epsilon+{\rm log}\left( \frac{\mu^2}{M^2} \right) \right){\rm tr}\{ F_{\mu\nu}F^{\mu\nu} \} \Bigg]
\\ \nonumber&&
+\int d^4x\Bigg[-\frac{ig^3}{(4\pi)^2}\frac{1}{90}\frac{1}{M^2}{\rm tr}\{ F_{\mu\nu}F^{\nu\sigma}F_\sigma\hspace{0.001cm}^\mu \}
\\ \nonumber &&
-\frac{g^2}{(4\pi)^2}\frac{1}{60}\frac{1}{M^2}{\rm tr}\{ D_\mu F^{\mu\nu}D^\sigma F_{\sigma\nu} \}\Bigg]+{\cal O}(1/M^4).
\end{eqnarray}
By taking $U_{\mu\nu}=-4igF_{\mu\nu}$ in Eq.(\ref{pgtexp}), we particularize this expression to the pure--gauge trace of the effective action $S^1$. By doing so and utilizing the expression for the scalar trace, we derive the effective Lagrangian expansion,
\begin{eqnarray}
{\cal L}^1&=&{\cal L}_{\rm YM}+\frac{g^2}{(4\pi)^2}\frac{31}{8}\sum_{m=1}^\infty\left[ \Delta_\epsilon+{\rm log}\left( \frac{R^2\mu^2}{m^2} \right)+\frac{2}{93} \right]{\rm tr}\{ F_{\mu\nu}F^{\mu\nu} \}
\\ \nonumber &&
-\frac{ig^3}{(4\pi)^2}\frac{281}{60}\sum_{m=1}^\infty\frac{R^2}{m^2}{\rm tr}\{ F_{\mu\nu}F^{\nu\sigma}F_\sigma\hspace{0.001cm}^\mu \}
-\frac{g^2}{(4\pi)^2}\frac{323}{120}\sum_{m=1}^\infty\frac{R^2}{m^2}{\rm tr}\{ D_\mu F^{\mu\nu}D^\sigma F_{\sigma\nu} \}
+{\cal O}(R^4).
\end{eqnarray}
Note that the only term involving UVd is the second one. In fact, this term also has discrete divergencies engendered by a KK infinite sum, and it introduces nondecoupling effects through the logarithm ${\rm log}(R\mu/m)$. However, as we established in the last chapter, such contributions are unobservable and hence innocuous because they can be absorbed through renormalization by the parameters of the low--energy theory. The KK infinite sums located in all other terms are Riemann $\zeta$--functions, which are finite and whose precise solutions are well known. As the second term involves all of the divergencies, of both continuous and discrete origins, generated by the KK excited modes at the one--loop level on light Green's functions, we conclude that {\it the one--loop level effects on low--energy physics are renormalizable, even though the extra--dimensional theory is known beforehand to be nonrenormalizable}. Recall that this conclusion was reached in the last chapter by following a different path. The renormalization proposed in the last chapter was achieved by means of the definitions
\begin{eqnarray}
A^{{\rm R}(0)a}_\mu&=&(1+L)^{\frac{1}{2}}A^{(0)a}_\mu,
\\ \nonumber \\
g_{\rm R}&=&(1+L)^{-\frac{1}{2}}g.
\end{eqnarray}
The powers of the coupling constants in each of the nonrenormalizable terms ensure that the usage of the renormalized fields and coupling constants do not modify the form of such terms, as all factors $(1+L)^{1/2}$ cancel. This means that we can just omit the unobservable effects without worrying about any other changes in the rest of the effective Lagrangian. With this in mind, we write the effective Lagrangian in the FtH gauge as
\begin{eqnarray}
\label{LeffFtH}
{\cal L}^1&=&{\cal L}_{\rm YM}-\frac{ig^3}{(4\pi)^2}\frac{281}{60}\sum_{m=1}^\infty\frac{R^2}{m^2}{\rm tr}\{ F_{\mu\nu}F^{\nu\sigma}F_\sigma\hspace{0.001cm}^\mu \}
\\ \nonumber &&
-\frac{g^2}{(4\pi)^2}\frac{323}{120}\sum_{m=1}^\infty\frac{R^2}{m^2}{\rm tr}\{ D_\mu F^{\mu\nu}D^\sigma F_{\sigma\nu} \}+{\cal O}(R^4).
\end{eqnarray}
This expansion has interesting features that deserve emphasis and discussion. First notice that it involves all independent nonrenormalizable terms of mass dimension six, which belong to the long list~\cite{effLagterms} of nonrenormalizable operators parametrizing physics beyond the SM. Of course, these operators are built of low--energy dynamic variables, which in this case are the KK zero modes $A^{(0)a}_\mu$, and they are governed by the four--dimensional Lorentz and gauge symmetries, which are the low--energy symmetries. A remarkable issue of the nonrenormalizable operators is the supression of the extra--dimensional effects provided by the compactification scale, $R^{-1}$, which also comes with the conclusion that the one--loop nonrenormalizable effects of the extra dimension are of decoupling nature, which can be appreciated by taking the limit of a very large compactification scale, $R^{-1}\to\infty$. This result was indeed expected because the fact that the low energy theory, ${\cal L}_{\rm YM}$, is renormalizable sets~\cite{W} the conditions for the decoupling theorem~\cite{AC} to hold, so that the high energy physics effects were, since the onset, expected to decouple. Note also the loop--factor $i/(4\pi)^2$ supressing the contributions from the nonrenormalizable terms. At the beginning of this section we argued in favor of the FtH gauge. I this sense, this effective expansion can be utilized to estimate effects of extra--dimensional physics on SM observables. For instance, one can, in the context of the electroweak SM, calculate the extra--dimensional contributions to the $S$, $T$, $U$ parameters~\cite{STUpar} by extracting the necessary tree--level two--point vertex functions from the ${\rm tr}\{ D_\mu F^{\mu\nu}D^\alpha F_{\alpha\nu} \}$ operator in Eq.(\ref{LeffFtH}) and find that the KK excited modes do not generate one--loop corrections to such parameters. One could also take the $WW\gamma$ and $WWZ$ tree--level vertices from both of the nonrenormalizable terms of Eq.(\ref{LeffFtH}) and compare the resulting form factors with those of Ref.~\cite{FMNRT}. The last part of the effective Lagrangian indicates that the next terms of the expansion shall be supressed by higher powers of the compactification radius.

Effective Lagrangians are a pragmatic and useful way to study physics beyond the SM, as they parametrize new--physics effects in a model--independent manner~\cite{W}. They are nonfundamental descriptions of nature that have a limited range of validity, so that as one approaches the characteristic cutoff of a given effective Lagrangian, the expansion becomes senseless and the full fundamental theory responsible of its effects enters as the genuine high--energy description of nature. For instance, the effective Lagrangian just derived in this section can be reliably used within a range of energies low enough below the compactification scale $R^{-1}$. One of the main features of the effective Lagrangian description is that these theories are built of low--energy dynamic variables and continuous symmetries, athough violations of discrete symmetries can appear. They can be used in two ways, depending on whether the fundamental theory is or is not known from the onset. If one departs with the precise knowledge of the high--energy theory, on can integrate out the heavy dynamic variables and calculate an effective Lagrangian involving nonrenormalizable terms of canonical dimensions higher than four, which will be accompanied by coefficients whose precise form shall be precisely known. This was the path followed in this thesis work. If one does not enjoy of accurate information about the nature of the fundamental theory, one can build by hand an effective Lagrangian by writing all permited nonrenormalizable operators to a certain order and multiply each one of them with a dimensionless unknown coefficient that parametrizes the impact of high energy physcs at the low--energy level. One can then calculate contributions of the nonrenormalizable operators to light observables and use experimental data to set bounds on the high energy physics that is the presumable origin of the effective Lagrangian. This was utilized, within the context of the electroweak SM, to set a bound~\cite{MQ,NTlic} on new--physics associated to CP--violating effects, which was done by investigating the contributions of the ${\rm tr}\{ F_{\mu\nu}F^{\nu\sigma}\tilde{F}_\sigma\hspace{0.001cm}^\mu \}$ operator, with $\tilde{F}_{\mu\nu}=(1/2)\varepsilon_{\mu\nu\rho\lambda}F^{\rho\lambda}$, to electric dipole moments of fermions induced at the one--loop level by the $W$ boson electric dipole moment. Another example of this can be found in Ref.~\cite{NRST}, where the contributions from the ${\rm tr}\{ F_{\mu\nu}F^{\nu\sigma}F_\sigma\hspace{0.001cm}^\mu \}$ low--energy invariant to the electromagnetic charge and anapole form factors were calculated, which was then used to set a bound on the neutrino charge radius. The effective Lagrangian ${\cal L}^1$, calculated in the present thesis work, involves only one physical parameter introduced by the extra--dimensional theory, namely, the compactification radius $R$. By vitrue of this, any KK contribution to low--energy physics shall contain such parameter, and when calculating an observable one can choose between using experimental limits on the size of the extra dimension to set bound on new--physics effects or take the experimental data concerning new--physics to establish bounds on $R$.

\section{The $R_\xi$ gauge}
So far, we have restricted ourselves to the FtH gauge, which, as we have argued, has interesting advantages from the physical and practical viewpoints. Nonetheless, a calculation in the general $R_\xi$ is a compelling part of any calculation involving viable sources of gauge dependence, as it occurs with the integration of KK excited gauge modes. For such a reason, in this section we shall derive the effective Lagrangian expansion in the $R_\xi$ gauge, which means that we shall keep the GF parameter $\xi$ unfixed. We start from the pure--gauge trace of the effective action $S^\xi$, Eq.(\ref{Seff1/2}), which in the general $R_\xi$ gauge reads
\begin{equation}
i\,{\rm Tr}\;{\rm log}\left[ g_{\mu\nu}(D^2+m_m^2)-\left( 1-\frac{1}{\xi} \right)D_\mu D_\nu-4igF_{\mu\nu} \right].
\end{equation}
In this case, the cross--derivative term $D_\mu D_\nu$ is still present, which greatly complexifies the calculation of the effective Lagrangian even if one tries to follow the procedure detailed in Appendix~\ref{app1}. Fortunately, there is a helpful trick that simplifies this calculation. It relies in noting that Lorentz covariance allows one to express the cross covariant derivative as
\begin{equation}
D_\mu D_\nu=\frac{1}{d}g_{\mu\nu}D^2-\frac{ig}{2}F_{\mu\nu},
\end{equation}
where the factor $1/d$ comes from the fact that we are working in $d$ dimensions. This result can be utilized to write the pure--gauge trace as
\begin{eqnarray}
&&i\,{\rm Tr}\;{\rm log}\left[ (D^2+m_m^2)g_{\mu\nu}-\left( 1-\frac{1}{\xi} \right)D_\mu D_\nu-4igF_{\mu\nu} \right]
\\ &&
=i\,{\rm Tr}\;{\rm log}\Bigg[ \left( D^2+\left( 1-\frac{\alpha}{d} \right)m_m^2 \right)g_{\mu\nu}-ig\left( 1-\frac{\alpha}{d} \right)\left( \frac{8-\alpha}{2} \right)F_{\mu\nu} \Bigg],
\end{eqnarray}
where $\alpha$ has been defined as
\begin{equation}
\label{alpha}
\alpha\equiv1-\frac{1}{\xi}.
\end{equation}
The gain of writing the pure--gauge trace in this manner is that now it has the same shape of Eq.(\ref{pgtexp}), so that we can directly employ this result. On the other hand, the gauge--dependent scalar trace has the same form than the one in the FtH gauge. Using all these results, we find that the corresponding effective Lagrangian is given by
\begin{eqnarray}
\label{LeffRxg}
{\cal L}^\xi&=&{\cal L}_{\rm YM}+\frac{ig^3}{(4\pi)^2}\frac{5\alpha^3-161\alpha^2+1528\alpha-4496}{60(4-\alpha)^2}\sum_{m=1}^\infty\frac{R^2}{m^2}\,{\rm tr}\{ F_{\mu\nu}F^{\nu\sigma}F_\sigma\hspace{0.001cm}^\mu \}
\\ \nonumber&&
-\frac{g^2}{(4\pi)^2}\frac{20\alpha^2-323\alpha+1292}{120(4-\alpha)}\sum_{m=1}^\infty\frac{R^2}{m^2}\,{\rm tr}\{ D_\mu F^{\mu\nu}D^\alpha F_{\alpha\nu} \}+{\cal O}(R^4),
\end{eqnarray}
where the unobservable terms, which are absorbed by renormalization, have been already omitted. It is evident that this expression carries gauge dependence through the factor $\alpha$, defined in Eq.(\ref{alpha}). Usually, although not necessarily~\cite{NTlic,Nmct,NTtedm}, loop calculations of Green's functions involving gauge fields into the loops lead to gauge--dependent results. It is only after all contributing Green's functions are summed together to construct an $S$--matrix element that such gauge dependence disappears. The nonrenormalizable terms of the effective Lagrangian ${\cal L}^\xi$ are equivalent to a sum of off--shell light Green's functions carrying one--loop contributions of the KK modes, that is, off--shell one--loop diagrams with KK zero modes as external legs and KK excited modes being the internal lines constituting the loops. From this perspective, it is natural that the effective Lagrangian ${\cal L}^\xi$ shows gauge dependence. It is important remarking that, as a consistency test, taking the particular gauge $\xi\to1$ in Eq.(\ref{LeffRxg}) leads to the recovery of the effective Lagrangian in the FtH gauge that we calculated in the previous section and exhibited in Eq.(\ref{LeffFtH}). The gauge dependence of an effective Lagrangian expansion produced by integrating out gauge heavy modes had not been explicitly proven and discussed in the literature before, so that the precise gauge--dependent expression derived in this work is a novel result. The possibility of having heavy massive gauge fields, even in frameworks different than extra dimensions, could in principle originate gauge dependence of effective theories. This, for instance, could be the case of the 331 model~\cite{331}, within which massive gauge fields are engendered after a symmetry breaking.

\section{The full effective theory}
The nonrenormalizable character of extra--dimensional models implies that there must be a more fundamental theory at a higher--energy scale, which can not be arbitrarily far from the compactification scale, provided that this higher--energy physics must control the divergencies that can not be absorbed by the extra--dimensional theory itself. The physics of the fundamental theory can be studied at lower energy by parametrizing its effects through a five--dimensional effective Lagrangian including operators of mass dimensions higher than five that are constituted by five--dimensional gauge fields ${\cal A}^a_M(x,y)$ and that are invariant under the ${\rm SU}_5(N)$ gauge transformations. The higher--than--five canonical dimension series should look like
\begin{equation}
\sum_N\beta_N\frac{g_5^{n_N}}{M_{\rm S}^{m_N}}{\cal O}^{\rm 5\,dim}_N({\cal A}^a_M),
\end{equation}
where the ${\cal O}^{\rm 5\,dim}_N$ are the operators of canonical dimension higher than five. The mass dimensions of each term are regulated by appropriate powers of the dimensionful coupling constant $g_5$ and the fundamental scale $M_{\rm S}$. On the other hand, the coefficients $\beta_N$, which quantify the contributions of the higher--energy physics, are dimensionless. This term must be added to the five--dimensional Yang--Mills theory ${\cal L}_{\rm YM}^{\rm 5\,dim}$ to complete the extra--dimensional effective description. Furthermore, these effects can be studied at the four--dimensional level by compactifying the extra dimension, then KK--expanding the covariant objects constituting the effective terms, and finally integrating out the extra dimension. All this process shall generate~\cite{UED1,NT2,FMNRT} four--dimensional nonrenormalizable operators of mass dimension higher than four whose dynamic variables and symmetries shall be those of the KK theory ${\cal L}_{\rm YM}^{\rm 4\,dim}$. As there are two high--energy scales, $R^{-1}$ and $M_{\rm S}$, the supression of the terms parametrizing the fundamental physics at the four--dimensional level si expected to involve powers of both of these scales. The full KK effective theory can be expressed as
\begin{eqnarray}
{\cal L}_{\rm eff}^\xi&=&{\cal L}_{\rm YM}^{\rm 4\,dim}(A^{(0)a}_\mu,A^{(m)a}_\mu,A^{(m)a}_5)+{\cal L}_{\rm GF}(A^{(0)a}_\mu,A^{(m)a}_\mu,A^{(m)a}_5;\xi)
\\ \nonumber&&
+{\cal L}_{\rm FPG}(A^{(0)a}_\mu,A^{(m)a}_\mu,A^{(m)a}_5,C^{(m)a},\bar{C}^{(m)a};\xi)+\sum_N\alpha_N\frac{R^{j_N}}{M_{\rm S}^{k_N}}{\cal O}^{\rm 4\,dim}_{N}(A^{(0)a}_\mu,A^{(m)a}_\mu,A^{(m)a}_5),
\end{eqnarray}
with the ${\cal O}^{\rm 4\,dim}_N$ representing the KK nonrenormalizable invariants associated to the fundamental physics beyond $M_{\rm S}$. On the other hand, the $\alpha_N$ are dimensionless and unknown coefficients that can be determined if the fundamental description is known or by employing experimental data. As the KK zero modes are the lightest fields, the less--supressed nonrenormalizable invariants shall be those involving exclusively these KK modes, so that, from here on, we will disregard those effective terms incorporating KK excited modes. Note that, by contrast with the ${\cal L}_{\rm YM}^{\rm 4\,dim}$, ${\cal L}_{\rm GF}$ and ${\cal L}_{\rm FPG}$ terms, the higher--than--four canonical dimension invariants ${\cal O}^{\rm 4\,dim}_N$ can contribute to low--energy physics since the tree level. After integrating out the KK excited modes of the Lagrangians ${\cal L}_{\rm YM}^{\rm 4\,dim}$, ${\cal L}_{\rm GF}$ and ${\cal L}_{\rm FPG}$ and ignoring the most--supressed ${\cal O}^{\rm 4\,dim}_N$ operators, the full effective Lagrangian ${\cal L}_{\rm eff}^\xi$ acquires the form
\begin{eqnarray}
\label{fullLEFF}
{\cal L}_{\rm eff}^\xi&=&{\cal L}_{\rm YM}+\frac{ig^3R^2}{5760}\frac{5\alpha^3-161\alpha^2+1529\alpha-4496}{(4-\alpha)^2}{\rm tr}\{ F_{\mu\nu}F^{\nu\sigma}F_\sigma\hspace{0.001cm}^\mu \}
\\ \nonumber&&
-\frac{g^2R^2}{11520}\frac{20\alpha^2-323\alpha+1292}{4-\alpha}{\rm tr}\{ D_\mu F^{\mu\nu}D^\alpha F_{\alpha\nu} \}+{\cal O}(R^4)+\sum_N\alpha_N\left( \frac{R}{M_{\rm S}} \right)^{r_N}{\cal O}^4_N(A^{(0)a}_\mu).
\end{eqnarray}

\subsection{Extra--dimensional physics versus fundamental description}
The full effective theory exhibited in Eq.(\ref{fullLEFF}) opens the possibility of comparing~\cite{NT2} in a rather economic manner the effects from the five--dimensional Yang--Mills theory with those originated by the fundamental theory beyond the $M_{\rm S}$ scale. Even though the fundamental physics contributes to low--energy physics since the tree level, while the extra--dimensional physics does it since the one--loop level, the supression provided by the $M_{\rm S}$ scale, which can be appreciated in Eq.(\ref{fullLEFF}), renders the extra--dimensional effects dominant over those from the fundamental theory. This was shown in Ref.~\cite{UED1} and phenomenologically illustrated in Ref.~\cite{FMNRT}. We commence by expressing the second and third terms of Eq.(\ref{fullLEFF}) in the FtH gauge as
\begin{eqnarray}
ig^3\kappa_{F}R^2\,{\rm tr}\{ F_{\mu\nu}F^{\nu\sigma}F_\sigma\hspace{0.001cm}^\mu \},
\\ \nonumber \\
g^2\kappa_{DF}R^2\,{\rm tr}\{ D_\mu F^{\mu\nu}D^\alpha F_{\alpha\nu} \},
\end{eqnarray}
where $\kappa_F$ and $\kappa_{DF}$ are dimensionless coefficients of order $10^{-2}$. On the other hand, the fundamental physics could engender the following operators of mass dimensions higher than five:
\begin{eqnarray}
\beta_{\cal F}\frac{ig_5^3}{M_{\rm S}}{\rm tr}\{ {\cal F}_{MN}{\cal F}^{NS}{\cal F}_S\hspace{0.001cm}^N \},
\\ \nonumber \\
\beta_{\cal DF}\frac{g_5^2}{M_{\rm S}}{\rm tr}\{ {\cal D}_M{\cal F}^{MN}{\cal D}^A{\cal F}_{AN} \}.
\end{eqnarray}
where ${\cal D}_M$ is the ${\rm SU}_5(N)$ covariant derivative in the adjoint representation of the gauge group. The dimensionless coefficients $\beta_{\cal F}$ and $\beta_{\cal DF}$ quantify the effects of the higher--energy fundamental description. For the construction of these five--dimensional invariants we have taken some care on writing a dimensionful coupling constant $g_5$ per each curvature. Such coupling constants regulate, altogether with the $M_{\rm S}$ scale, the mass dimensions of these terms. By compactifying the extra dimension and integrating it out, these higher--than--five canonical dimension operators produce the four--dimensional nonrenormalizable low--energy invariants
\begin{eqnarray}
\int_0^{\pi R}dy\left(\beta_{\cal F}\frac{ig_5^3}{M_{\rm S}}{\rm tr}\{ {\cal F}_{MN}{\cal F}^{NS}{\cal F}_S\hspace{0.001cm}^N \}\right)=\beta^{(0)}_{\cal F}ig^3\frac{R}{M_{\rm S}}{\rm tr}\{ F_{\mu\nu}F^{\nu\sigma}F_\sigma\hspace{0.001cm}^\mu \}+\cdots ,
\\ \nonumber \\
\int_0^{\pi R}dy\left(\beta_{\cal DF}\frac{g_5^2}{M_{\rm S}}{\rm tr}\{ {\cal D}_M{\cal F}^{MN}{\cal D}^A{\cal F}_{AN} \}\right)
=\beta^{(0)}_{\cal DF}g^2\frac{R}{M_{\rm S}}{\rm tr}\{ D_\mu F^{\mu\nu}D^\alpha F_{\alpha\nu} \}+\cdots,
\end{eqnarray}
We have defined the dimensionless coefficients $\beta^{(0)}_{\cal F}=\pi\beta_{\cal F}$ and $\beta_{\cal DF}^{(0)}=\pi\beta_{\cal DF}$, which parametrize the impact of the $M_{\rm S}$--scale physics on the low--energy standard theory. We have not explicitly written all other four--dimensional nonrenormalizable terms generated by the higher--energy description, as for this comparison only the very first terms, which are exclusively consituted by zero modes, are necessary. It is important emphasizing the presence of the supression factor $R/M_{\rm S}$ in these terms. The five--dimensional theory becomes~\cite{DDG1,UED1} nonperturbative at some high--energy scale, which could be identified as $M_{\rm S}$. Assuming that the extra--dimensional description is still perturbative up to such scale leads~\cite{UED1,UED2} to the estimation $M_{\rm S}R\sim30$. With this in mind, and taking $\beta^{(0)}_{\cal F}\sim\kappa_F$ and $\beta_{\cal DF}\sim\kappa_{DF}$, it is a straightforward matter to estimate that {\it the contributions of the fundamental physics are about 3\% of those produced by the extra--dimensional theory}. The domination of the extra--dimensional physics over the beyond--$M_{\rm S}$ scale theory is an outcome of the extra supression introduced by the $M_{\rm S}$ scale on the low--energy parametrization of the fundamental description.

\chapter{Conclusions}
\label{conclusions}
In the present thesis work we have been concerned with an extra--dimensional version of the Yang--Mills theory. We have assumed the existence of one extra dimension, which we supposed to be universal. This means that all of the fields of the theory are allowed to propagate in the extra dimension. We compactified the extra dimension on the orifold $S^1/Z_2$, which set the conditions to expand the five--dimensional gauge fields in Kaluza--Klein towers with well--defined parity with respect to the extra dimension. A remarkable result of this work was the observation that the objects to KK expand are the curvatures instead of the gauge fields, for the curvatures are covariant objects whose expansion allows the low--energy four--dimensional theory to inherit a rich gauge--symmetry structure. The KK theory obtained after compactifying the extra dimension and then integrating it out has remarkable features. The nature of the gauge symmetry governing the four--dimensional description is particularly appealing, as the compactification of the extra dimension produces an infinite number of gauge transformations that can be divided into two types. One of them is constituted by the standard gauge transformations, which correspond to the usual low--energy ${\rm SU}_4(N)$ gauge variations. On the other hand, a new sort of complicated gauge transformations, which we called the nonstandard gauge transformations, emerges. According to the structure of both of these sets of gauge transformations, the KK zero modes $A^{(0)a}_\mu$ are gauge fields under the SGT, the KK excited modes $A^{(m)a}_\mu$ are gauge fields under the NSGT, and the scalar fields $A^{(m)a}_5$ are pseudo--Goldstone bosons that can be removed from the theory by choosing a particular gauge with respect to the NSGT. The quantization of the KK theory was another main issue of this thesis work, and it is worth emhasizing that it had not been consistently discussed before in the literature. The quantization was performed on the grounds of the BRST formulation. An interesting feature of the KK theory that we derived is that the existence of two independent sets of gauge transformations, defined by independent gauge parameters, opens the possibility of quantizing the zero modes and the excited ones independently of each other. We utilized this fact to quantize the KK excitations and then we could quantize the low--energy theory and prove that the one--loop level contributions of the KK excited modes to low--energy physics are renormalizable, which is very important, as it ensures that the very first contributions of the extra--dimensional physics to light observables can be calculated without worrying about ambiguities introduced by an infinite number of parameters absorbing divergencies. As the two quantization steps were independent of each other, we were able to use different schemes to quantize the zero and the excited modes. We paid especial attention to the quantization of the KK excited modes, for which we introduced a profitable GF procedure that allowed us to preserve gauge invariance under the SGT and also introduced significative simplifications for further developments of the paper. This GF procedure is based on a proposal made in the literature for the 331 model, which we found to have some similarities with the KK theory that we analyzed through this thesis work. The quantization of the KK excited modes of the KK theory produced the quantum version of the Lagrangian, which comprises the KK theory, the GF term, and the most general FPG sector. The knowledge of the precise form of the KK quantum Lagrangian allowed us to integrate out the KK excited modes, which we did by adjusting a method proposed some years ago by M. Bilenky and A. Santamaria. Such an integration comprehended the KK excited gauge fields, the pseudo--Goldstone bosons and the KK excited modes of the ghost fields. This integration is novel, as the integration of heavy gauge fields to obtain an effective Lagrangian is a rahter ignored issue. Moreover, it brings interesting possibilities, as the gauge--dependent behavior of the gauge modes perseveres to the point that the resulting effective Lagrangian expansion is gauge dependent. This feature is indeed consistent, for the effective Lagrangian so obtained is equivalent to a sum of one--loop light Green's functions with light external legs, but KK excited modes exclusively filling the loops. During the development of the calculation, we were able to elegantly and explicitly show that, under the SGT--covariant GF procedure that we introduced for the KK excitations, the contributions of the pseudo--Goldstone bosons are minus twice the ones generated by the ghost fields. We were also able to confirm, through the derivation of the effective Lagrangian, that the contributions of the KK excited modes at the one--loop level are renormalizable, as the divergencies generated by the KK excited modes are absorbed by the parameters of the low--energy theory. Our expression of the gauge--dependent effective Lagrangian involves four--dimensional nonrenormalizable operators of mass dimensions six, which are governed by the light symmetries, Lorentz and ${\rm SU}_4(N)$, and whose dynamic variables are only KK zero modes. Finally, the effective Lagrangian that we derived allowed us to perform in a simple way a comparison among the impact, at low--energy, of the extra--dimensional effects and the ones produced by the fundamental theory, lying beyond extra dimensions. We found, within the Feynman--`t Hooft gauge, that the effects of the extra--dimensional physics are dominant, for the contributions from the fundamental description represent around $3\%$ of their impact. This behavior is provided by an extra supression introduced by the fundamental scale, $M_{\rm S}$.

\appendix

\chapter{A gauge determinant}
\label{app1}
The functional integration of heavy fields in a theory describing physics at certain high energy scale leads to determinants that concern all of the light degrees of freedom, both internal and external. The determinants so obtained can be transformed into traces over all of such degrees of freedom. In this appendix, we adjust the method presented in Ref.~\cite{BS} to calculate a trace carrying the contributions of heavy gauge bosons and obtain a low--energy expansion up to canonical--dimension--six non--renormalizable operators. Consider the general trace
\begin{equation}
i\hspace{0.05cm}{\rm Tr}\hspace{0.05cm}{\rm log}\left[ g_{\mu\nu}(D^2+M^2)+U_{\mu\nu}(x) \right]\equiv\int d^4x\hspace{0.05cm}{\cal L}_{\rm 1-loop}(x)
\label{Trdef}
\end{equation}
where $D_\mu$ is the covariant derivative for the ${\rm SU}(N)$ gauge group and $U_{\mu\nu}(x)$ is a space--time dependent matrix that we suppose to be arbitrary. The covariant derivative is given by
\begin{equation}
\begin{array}{lr}
D_\mu=\partial_\mu+G_\mu,&\hspace{0.5cm}G_\mu=-igT^aG^a_\mu,
\end{array}
\end{equation}
with $T^a$ representing the generators of the gauge group and $G^a_\mu$ standing for the gauge fields. The curvature, which we denote by $G_{\mu\nu}^a$, is defined in terms of the covariant derivative as
\begin{equation}
\begin{array}{lr}
G_{\mu\nu}=\left[ D_\mu,D_\nu \right],&\hspace{0.5cm}G_{\mu\nu}=-igT^aG^a_{\mu\nu}.
\label{curvadef}
\end{array}
\end{equation}
The trace operation in Eq.(\ref{Trdef}) acts on the points of the space--time, which are the external degrees of freedom. It also affects the internal degrees of freedom, which in this case are determined by the gauge and Lorentz groups. In the following, the symbol ``Tr" shall refer to a trace over both the external and the internal degrees of freedom, while ``tr" shall indicate a trace over internal degrees of freedom, exclusively. As divergencies shall appear, below, they must be appropriately regularized. We follow the dimensional regularization approach, for which we work, from here on, in $d$ dimensions. Note that the argument of the trace is non--local, so that, up to this point, performing this operation makes no sense. To obtain a local expression, one can first perform the trace over the space--time coordinates, which, for a general operator ${\cal O}$, should be understood as
\begin{equation}
{\rm Tr}\left\{ {\cal O} \right\}=\int d^dx\hspace{0.1cm}{\rm tr}\hspace{0.05cm}\langle x|{\cal O}|x\rangle=\int d^dx\hspace{0.05cm}d^d\tilde{p}\hspace{0.1cm}{\rm tr}\left\{ \langle x | {\cal O} | p \rangle \langle p | x \rangle \right\}
\end{equation}
where a completeness relation has been inserted and we have defined
\begin{equation}
d^d\tilde{p}=\mu^{(4-d)/2}\frac{d^dp}{(2\pi)^d},
\end{equation}
so that $\mu$ is a factor introduced to appropriately correct dimensions. For a general quantum state, $|\alpha\rangle$,
\begin{equation}
\langle x | {\cal O} | \alpha\rangle={\cal O}_x\langle x|\alpha\rangle={\cal O}_x\hspace{0.05cm}\alpha(x),
\end{equation}
with ${\cal O}_x$ standing for the operator ${\cal O}$ in the representation of positions. With this in mind, note that
\begin{equation}
{\rm Tr}\{ {\cal O} \}=\int d^dx\hspace{0.05cm}d^d\tilde{p}\hspace{0.1cm}{\rm tr}\left\{ e^{ip\cdot x}{\cal O}_xe^{-ip\cdot x} \right\}
\label{Tredf}.
\end{equation}
By defining $\Pi_\mu\equiv iD_\mu$ and then applying the general result shown in Eq.(\ref{Tredf}), along with the operator identity $e^{ip\cdot x}f(\Pi)e^{-ip\cdot x}=f(\Pi+p)$, to the gauge trace, Eq.(\ref{Trdef}), one obtains
\begin{eqnarray}
i\hspace{0.05cm}{\rm Tr}\hspace{0.05cm}{\rm log}\left[ g_{\mu\nu}(D^2+M^2)+U_{\mu\nu} \right]&=&i\hspace{0.05cm}{\rm Tr}\hspace{0.05cm}{\rm log}\left[ g_{\mu\nu}\left( -\Pi^2+M^2 \right)+U_{\mu\nu} \right]
\\ \nonumber&=&
 i\int d^dx\hspace{0.05cm}d^d\tilde{p}\hspace{0.1cm}{\rm tr}\Bigg\{ {\rm log}\left[ -p^2+M^2 \right]
\\ \nonumber &&
+{\rm log}\left[ g_{\mu\nu}+\frac{(\Pi^2+2\Pi\cdot p)g_{\mu\nu}-U_{\mu\nu}}{p^2-M^2} \right] \Bigg\}{\bf 1}
\end{eqnarray}
In this expression, the logarithm operators act on the identity, which we denote by ``{\bf 1}". The first term of the argument of the trace in the second line of the last expression contains no fields, and hence contributes only to the vacuum energy density. Thus, we drop it in what follows and conserve only the second term, so that the gauge trace is expanded as
\begin{equation}
i\hspace{0.05cm}{\rm Tr}\hspace{0.05cm}{\rm log}\left[ g_{\mu\nu}(D^2+M^2)+U_{\mu\nu} \right]=i\int d^dx\hspace{0.05cm}d^d\tilde{p}\hspace{0.1cm}{\rm tr}\sum_{k=1}^\infty\frac{(-1)^{k+1}}{k}\frac{\left[ (\Pi^2+2\Pi\cdot p)\delta^\mu\hspace{0.001cm}_\nu-U^\mu\hspace{0.001cm}_\nu \right]^k}{\left[ p^2-M^2 \right]^k}\hspace{0.05cm}{\bf 1}.
\label{Trexpansion}
\end{equation}
The argument of the gauge trace, written in this form, is local, so that the trace over the internal degrees of freedom can be taken in each term of the series.

From the local form of Eq.(\ref{Trexpansion}), one can appreciate that the calculation of the momentum integrals shall provide an expansion of non--renormalizable operators, each one multiplied by a power of $M$. In other words, by comparing Eqs.(\ref{Trdef}) and (\ref{Trexpansion}), one can extract the Lagrangian
\begin{equation}
{\cal L}_{\rm 1-loop}=\sum_{k=1}^\infty\frac{(-1)^{k+1}}{k}\hspace{0.05cm}{\rm tr}\left\{ i\int d^d\tilde{p}\frac{\left[ (\Pi^2+2\Pi\cdot p)\delta^\mu\hspace{0.001cm}_\nu-U^\mu\hspace{0.001cm}_\nu \right]^k}{\left[ p^2-M^2 \right]^k} {\bf 1} \right\},
\label{appL1loopexp}
\end{equation}
then solve the loop integrals term by term in the series, and finally write ${\cal L}_{\rm 1-loop}$ as
\begin{equation}
{\cal L}_{\rm 1-loop}=d\sum_{k=1}^\infty\frac{c_k}{M^{2k-4}}{\cal O}_k,
\label{Lexpgenform}
\end{equation}
with ${\cal O}_k$ representing a linear combination of traces of gauge invariant operators of canonical dimension $2k$, built of the gauge fields $G^a_\mu$, the matrix $U_{\mu\nu}$, and the ${\rm SU}(N)$ covariant derivative. In the ${\cal L}_{\rm 1-loop}$ expansion shown in Eq.(\ref{Lexpgenform}), there is a global factor, $d$, which is expected because the gauge trace, Eq.(\ref{Trdef}), was produced by the integration of vector fields, which are constituted by $d$ scalar fields, and the sum of all contributions produces this global factor. A convenient normalization of this expansion fixes the $c_k$ coefficients as
\begin{equation}
c_k=\frac{1}{(4\pi)^2}\left( \frac{M^2}{4\pi\mu^2} \right)^{d/2-2}\Gamma\left( k-\frac{d}{2} \right).
\label{normalization}
\end{equation}
As already commented, the ${\cal O}_k$ are combinations of traces of dimension--2$k$ operators, so that, in general, they should have the form
\begin{equation}
{\cal O}_k=\sum_j a_{k,j}\hspace{0.05cm}{\cal O}_{k,j}.
\label{tracescombination}
\end{equation}
For the first values of $k$ ($=1,2,3$), we shall employ the following sets as bases:
\begin{eqnarray}
k=1:&\hspace{0.5cm}&\bigg({\rm tr}\{ U^\mu\hspace{0.001cm}_\mu \}\bigg)
\label{bk1}
\\ \nonumber \\
k=2:&\hspace{0.5cm}&\bigg( {\rm tr}\{ U_{\mu\nu} U^{\nu\mu} \},d\hspace{0.05cm}{\rm tr}\{ G_{\mu\nu}G^{\mu\nu} \} \bigg)
\label{bk2}
\\ \nonumber \\
\label{bk3}
k=3:&\hspace{0.5cm}&\bigg( {\rm tr}\{ U_{\mu\nu}U^{\nu\sigma}U_\sigma\hspace{0.001cm}^\mu \},{\rm tr}\{ D_\mu U^{\mu\nu}D^\sigma U_{\sigma\nu} \},{\rm tr}\{ G_{\mu\nu}U^{\nu\sigma}G_\sigma\hspace{0.001cm}^\mu \},
\\ \nonumber&&
d\hspace{0.05cm}{\rm tr}\{ D_\mu G^{\mu\nu}D^\sigma G_{\sigma\nu} \},
d\hspace{0.05cm}{\rm tr}\{ G_{\mu\nu}G^{\nu\sigma}G_\sigma\hspace{0.001cm}^\mu \} \bigg)
\end{eqnarray}
The method proposed in Ref.~\cite{BS} relies on the fact that Eqs.(\ref{appL1loopexp}), (\ref{Lexpgenform}) and (\ref{tracescombination}) are valid for any field configuration. In fact, the $a_{k,j}$ coefficients are just numbers, independent of the field configuration, so that if one is able to determine them within a specific choice, the coefficients corresponding to the bases exhibited in Eqs.(\ref{bk1}), (\ref{bk2}) and (\ref{bk3}) can be obtained.  An appropriate election is the field configuration such that
\begin{equation}
\begin{array}{lll}
E_\mu\equiv G_\mu,&\hspace{0.1cm}\partial_\mu E_\nu=0,&\hspace{0.01cm}U^\mu\hspace{0.001cm}_\nu=-\delta^\mu\hspace{0.001cm}_\nu E^2,
\label{scconditions1}
\end{array}
\end{equation}
which imply that
\begin{equation}
\begin{array}{ll}
G_{\mu\nu}=[E_\mu,E_\nu],&\hspace{0.1cm}D_\mu{\cal M}=[E_\mu,{\cal M}],
\label{scconditions2}
\end{array}
\end{equation}
where ${\cal M}$ represents any matrix valued function of $E_\mu$ and $U^\mu\hspace{0.001cm}_\nu$. In these very particular circumstances, the numerator of the ${\cal L}_{\rm 1-loop}$ expansion, Eq.(\ref{appL1loopexp}), is greatly simplified, for each term of the series can be expressed as
\begin{equation}
\left[ (\Pi^2+2\Pi\cdot p)\delta^\mu\hspace{0.001cm}_\nu -U^\mu\hspace{0.001cm}_\nu \right]^k{\bf 1}=\delta^\mu\hspace{0.001cm}_\nu(2iE\cdot p)^k,
\end{equation}
which considerably reduces the procedure of calculating the momentum integrals:
\begin{eqnarray}
\label{steps}
{\cal L}_{\rm 1-loop}(E_\mu)&=&\sum_{k=1}^\infty\frac{(-1)^{k+1}}{k}\hspace{0.05cm}{\rm tr}\left\{ \int d^d\tilde{p}\hspace{0.05cm}\frac{\delta^\mu\hspace{0.001cm}_\nu (2iE\cdot p)^k}{(p^2-M^2)^k} \right\}
\\ \nonumber 
&=&d\hspace{0.05cm}\sum_{k=1}^\infty\frac{(-1)^{k+1}4^k}{2k}\hspace{0.05cm}i\int d^d\tilde{p}\hspace{0.05cm}\frac{p^{\mu_1}p^{\mu_2}\ldots p^{\mu_{2k}}}{(p^2-M^2)^{2k}}\hspace{0.05cm}{\rm tr}\left\{ E_{\mu_1}\ldots E_{\mu_{2k}} \right\}
\\ \nonumber
&=&d\hspace{0.05cm}\sum_{k=1}^\infty\frac{1}{M^{2k-4}}\frac{1}{(4\pi)^2}\left( \frac{M^2}{4\pi\mu^2} \right)^{d/2-2}\Gamma\left( k-\frac{d}{2} \right)\frac{2^k}{(2k)!}\hspace{0.05cm}{\rm tr}\left\{ {\cal S}_{2k}(E_\mu) \right\}
\\ \nonumber
&=&d\hspace{0.05cm}\sum_{k=1}^\infty\frac{c_k}{M^{2k-4}}\frac{2^k}{(2k)!}\hspace{0.05cm}{\rm tr}\left\{ {\cal S}_{2k}(E_\mu) \right\} ,
\end{eqnarray}
In passing from the first to the second line in Eq.(\ref{steps}), we took the trace over the Lorentz indices, which gave rise to the $d$ factor (recall that we are working in $d$ dimensions!). We then considered the fact that any loop integral with an odd number of momentum factors vanishes. From the second to the third line, we solved the momentum integrals by utilizing the result
\begin{equation}
i\int d^d\tilde{p}\hspace{0.1cm}\frac{p^{\mu_1}p^{\mu_2}\cdots p^{\mu_{2k}}}{(p^2-M^2)^{2k}}=(-1)^{k+1}\left( \frac{M^2}{4\pi\mu^2} \right)^{d/2-2}\frac{1}{(4\pi)^2}\frac{1}{M^{2k-4}}\frac{\Gamma(k-d/2)}{2^k\Gamma(2k)}S_{k}^{\mu_1 \mu_2\ldots\mu_{2k}}.
\end{equation}
In this expression, $S_{k}^{\mu_1\ldots\mu_{2k}}$ is a totally symmetric tensor built of the sum of all the products of $k$ metric tensors involving all the possible permutations of Lorentz indices. For instance, $S_2^{\mu_1\ldots\mu_4}=g^{\mu_1\mu_2}g^{\mu_3\mu_4}+g^{\mu_1\mu_3}g^{\mu_2\mu_4}+g^{\mu_1\mu_4}g^{\mu_2\mu_3}$. Also, we have employed the definition
\begin{equation}
{\cal S}_{2k}(E_\mu)\equiv S_k^{\mu_1\ldots\mu_{2k}}E_{\mu_1}\ldots E_{\mu_{2k}},
\end{equation}
so that ${\cal S}_{2k}(E_\mu)$ is the sum of all possible permutations of products of $2k$ $E_\mu$ fields in which all of such fields are Lorentz--contracted. For example, ${\cal S}_4(E_\mu)=(E^2)^2+E_\mu E_\nu E^\mu E^\nu+E_\mu E^2E^\mu$. Finally, from the third to the fourth line, we have used the normalization of the $c_k$, Eq.(\ref{normalization}). By comparing the last line of Eq.(\ref{steps}) with the general expansion exhibited in Eq.(\ref{Lexpgenform}), one can identify
\begin{equation}
{\cal O}^{\rm S}_k=\frac{2^k}{(2k)!}\hspace{0.05cm}{\rm tr}\{ {\cal S}_{2k}(E_\mu) \},
\label{expsc}
\end{equation}
with the superscript "S" indicating that we are working in the specific field configuration. Within this especial configuration, the ${\cal O}_k$, can be expanded as
\begin{equation}
{\cal O}^{\rm S}_k=\sum_ia^{\rm S}_{k,j}{\cal O}^{\rm S}_{k,j}.
\end{equation}
In this context, an appropriate set of bases of traces ${\cal O}^{\rm S}_{k,j}$ is
\begin{eqnarray}
k=1:&\hspace{0.5cm}&\bigg( {\rm tr}\{ E^2 \} \bigg)
\\ \nonumber \\
k=2:&\hspace{0.5cm}&\bigg( {\rm tr}\{ E_\mu E_\nu E^\mu E^\nu \},{\rm tr}\{ (E^2)^2 \} \bigg)
\\ \nonumber \\
k=3:&\hspace{0.5cm}&\bigg( {\rm tr}\{ E_\mu E_\nu E_\sigma E^\mu E^\nu E^\sigma \},{\rm tr}\{ (E^2)^3 \},{\rm tr}\{ E^2E_\mu E^2E^\mu \},{\rm tr}\{ E_\mu E_\nu E^\mu E_\sigma E^\nu E^\sigma \},
\\ \nonumber &&
{\rm tr}\{ E^2E_\mu E_\nu E^\mu E^\nu \} \bigg).
\end{eqnarray}
By employing Eq.(\ref{expsc}), one can straightforwardly obtain the ${\cal O}_k$ combinations in the especial configuration:
\begin{eqnarray}
{\cal O}^{\rm S}_1&=&{\rm tr}\{ E^2 \},
\label{sccombO1}
\\ \nonumber \\
{\cal O}^{\rm S}_2&=&\frac{1}{3}{\rm tr}\{ (E^2)^2 \}+\frac{1}{6}{\rm tr}\{ E_\mu E_\nu E^\mu E^\nu \},
\label{sccombO2}
\\ \nonumber \\
\label{sccombO3}
{\cal O}^{\rm S}_3&=&\frac{1}{90}{\rm tr}\{ E_\mu E_\nu E_\sigma E^\mu E^\nu E^\sigma \}+\frac{1}{45}{\rm tr}\{ (E^2)^3 \}+\frac{1}{30}{\rm tr}\{ E^2E_\mu E^2E^\mu \}
\\ \nonumber
&&+\frac{1}{30}{\rm tr}\{ E_\mu E_\nu E^\mu E_\sigma E^\nu E^\sigma \}
+\frac{1}{15}{\rm tr}\{ E^2E_\mu E_\nu E^\mu E^\nu \}.
\end{eqnarray}
One can also write the combinations ${\cal O}_k$ in terms of the bases in Eqs.(\ref{bk1}), (\ref{bk2}) and (\ref{bk3}), according to the general expression shown in Eq.(\ref{tracescombination}), and specialize the results to the special configuration, which was defined through Eqs.(\ref{scconditions1}) and (\ref{scconditions2}), as
\begin{equation}
{\cal O}^{\rm S}_k=\left.\sum_ia_{k,j}{\cal O}_{k,j}\right|_{\rm S}.
\end{equation}
So far, the $a_{k,j}$ coefficients remain unknown, but by equalizing the resulting expressions to Eqs.(\ref{sccombO1}), (\ref{sccombO2}) and (\ref{sccombO3}) as
\begin{equation}
{\cal O}^{\rm S}_k=\left.\sum_ia_{k,j}{\cal O}_{k,j}\right|_{\rm S}=\sum_ia_{k,j}^{\rm S}{\cal O}_{k,j}^{\rm S},
\end{equation}
one can determine such coefficients, which are independent of the configuration. We find
\begin{eqnarray}
{\cal O}_1&=&-\frac{1}{d}{\rm tr}\{  U^\mu\hspace{0.001cm}_\mu \},
\\ \nonumber \\
{\cal O}_2&=&\frac{1}{2d}{\rm tr}\{ U_{\mu\nu}U^{\nu\mu} \}+\frac{1}{12}{\rm tr}\{ G_{\mu\nu}G^{\mu\nu} \},
\\ \nonumber \\
{\cal O}_3&=&-\frac{1}{6d}{\rm tr}\{ U_{\mu\nu}U^{\nu\sigma}U_\sigma\hspace{0.001cm}^\mu \}+\frac{1}{12}{\rm tr}\{ D_\mu U^{\mu\nu}D^\sigma U_{\sigma\nu} \}+\frac{1}{12}{\rm tr}\{ G_{\mu\nu}U^{\nu\sigma}G_\sigma\hspace{0.001cm}^\mu \}
\\ \nonumber
&&+\frac{1}{60}{\rm tr}\{ D_\mu G^{\mu\nu}D^\sigma G_{\sigma\nu} \}
-\frac{1}{90}{\rm tr}\{ G_{\mu\nu}G^{\nu\sigma}G_\sigma\hspace{0.001cm}^\mu \}.
\end{eqnarray}
By inserting these results into Eq.(\ref{Lexpgenform}) along with the $c_k$ coefficients given by Eq.(\ref{normalization}), we obtain the following low--energy expansion,
\begin{eqnarray}
\label{finalloopexpansion}
{\cal L}_{\rm 1-loop}&=&\frac{1}{(4\pi)^2}M^2\left( \Delta_\epsilon
+{\rm log}\left( \frac{\mu^2}{M^2} \right)+1 \right){\rm tr}\{ U^\mu\hspace{0.001cm}_\mu \}
\\ \nonumber &&
+\frac{1}{(4\pi)^2}\frac{1}{2}\left( \Delta_\epsilon+{\rm log}\left( \frac{\mu^2}{M^2} \right) \right){\rm tr}\{ U_{\mu\nu}U^{\mu\nu} \}
\\ \nonumber&&
+\frac{1}{(4\pi)^2}\frac{1}{3}\left( \Delta_\epsilon+{\rm log}\left( \frac{\mu^2}{M^2}-\frac{1}{2} \right) \right){\rm tr}\{ G_{\mu\nu}G^{\mu\nu} \}-\frac{1}{(4\pi)^2}\frac{1}{M^2}\frac{1}{6}{\rm tr}\{ U_{\mu\nu}U^{\nu\sigma}U_\sigma\hspace{0.001cm}^\mu \}
\\ \nonumber&&
+\frac{1}{(4\pi)^2}\frac{1}{M^2}\frac{1}{3}{\rm tr}\{ D_\mu U^{\mu\nu}D^\sigma U_{\sigma\nu} \}+\frac{1}{(4\pi)^2}\frac{1}{M^2}\frac{1}{3}{\rm tr}\{ G_{\mu\nu}U^{\nu\sigma}G_\sigma\hspace{0.001cm}^\mu \}
\\ \nonumber&&
+\frac{1}{(4\pi)^2}\frac{1}{M^2}\frac{1}{15}{\rm tr}\{ D_\mu G^{\mu\nu}D^\sigma G_{\sigma\nu} \}-\frac{1}{(4\pi)^2}\frac{1}{M^2}\frac{2}{45}{\rm tr}\{ G_{\mu\nu}G^{\nu\sigma}G_\sigma\hspace{0.001cm}^\mu \}.
\end{eqnarray}
with
\begin{eqnarray}
\Delta_\epsilon=\frac{1}{\epsilon}-\gamma_E+{\rm log}(4\pi),&\hspace{0.5cm}&\epsilon=\frac{4-d}{2}.
\end{eqnarray}
As a final remark, note that, by virtue of Eq.(\ref{curvadef}), one should perform the change $G_{\mu\nu}\rightarrow-ig\hspace{0.01cm}G_{\mu\nu}$ in Eq.(\ref{finalloopexpansion}) in order to be in agreement with the standard notation.

\end{document}